\documentclass[draft]{agujournal2019}
\usepackage{url} 
\usepackage{lineno}
\usepackage[inline]{trackchanges} 

\usepackage{amsmath,amsfonts,amssymb,amsthm}

\usepackage{soul}
\usepackage{tikz}

\newcounter{cnt_cr}

\draftfalse

\journalname{Journal of Advances in Modeling Earth Systems (JAMES)}

\begin{document}

%
%


\title{A Constrained Spectral Approximation of Subgrid-Scale Orography on Unstructured Grids}

%
%




\authors{
Ray Chew\affil{1}, Stamen Dolaptchiev\affil{1}, Maja-Sophie Wedel\affil{1}, Ulrich Achatz\affil{1}
}

\affiliation{1}{Institute for Atmospheric and Environmental Sciences, Goethe University Frankfurt, Germany}





\correspondingauthor{Ray Chew}{chew@iau.uni-frankfurt.de}




\begin{keypoints}
\item We introduce a novel scale-aware spectral approximation method for subgrid-scale orographic data on non-quadrilateral geodesic grids
\item The method yields physically sound results and achieves reasonable error scores when approximating real-world orographic spectra
\item The method features over two orders of magnitude compression in complexity and has potential applications in generic spectral analyses
\end{keypoints}

%
%

%
%


\begin{abstract}
The representation of subgrid-scale orography is a challenge in the physical parameterization of orographic gravity-wave sources in weather forecasting. A significant hurdle is encoding as much physical information with as simple a representation as possible. Other issues include scale awareness, i.e., the orographic representation has to change according to the grid cell size and usability on unstructured geodesic grids with non-quadrilateral grid cells. This work introduces a novel spectral analysis method approximating a scale-aware spectrum of subgrid-scale orography on unstructured geodesic grids. The dimension of the physical orographic data is reduced by more than two orders of magnitude in its spectral representation. Simultaneously, the power of the approximated spectrum is close to the physical value. The method is based on well-known least-squares spectral analyses. However, it is robust to the choice of the free parameters, and tuning the algorithm is generally unnecessary. Numerical experiments involving an idealized setup show that this novel spectral analysis performs significantly better than a straightforward least-squares spectral analysis in representing the physical energy of a spectrum. Studies involving real-world topographic data are conducted, and reasonable error scores within $\pm 10\%$ error relative to the maximum physical quantity of interest are achieved across different grid sizes and background wind speeds. The deterministic behavior of the method is investigated along with its principal capabilities and potential biases, and it is shown that the error scores can be iteratively improved if an optimization target is known. Discussions on the method's limitations and broader applicability conclude this work.
\end{abstract}

\section*{Plain Language Summary}
Wind flow over terrain has wide-ranging influences on atmospheric processes. Meteorologists want to include this effect in their weather forecasts but encounter computational limitations. For global weather forecasting, terrain features are relatively small and usually not explicitly represented in forecast models. An ongoing research question is how best to represent these small-scale features in forecast models. However, a few difficulties arise:
\begin{enumerate}
    \item We want to encode as much information about the terrain with as simple a representation as possible.
    \item Some forecast models represent the globe with an icosahedron. Therefore, terrain information must be encoded within the triangular or hexagonal cells.
    \item The information encoded has to change if the size of the triangles or hexagons changes.
\end{enumerate}
In this work, we present a novel method to overcome the three difficulties mentioned above. We can compress terrain information from over 50\,000 to below 100 data points that can be used for climate simulations. When the relative effect of the encoded terrain on atmospheric processes is considered, the method has good accuracy despite the severe constraints and data compression, with a 10\% error bound. Apart from representing terrain in weather forecasting, this method has potential applications for generic data analysis.

\section{Introduction}
\label{sec:intro}
The impact of gravity waves on atmospheric flow is well known and has been extensively studied \cite<see, e.g.,>{achatz2024atmospheric,fritts2003gravity,kim2003overview,alexander2010recent}. Gravity waves arise from the perturbation of air parcels and the subsequent restoration via gravity or buoyancy. A significant source of gravity waves is the airflow over orography such as mountains and hills, and the modeling of orographic gravity waves presents an ongoing challenge for numerical weather prediction \cite<NWP;>{elvidge2019uncertainty, sandu2019impacts}. Direct simulation of resolved gravity waves requires a fine grid on the dynamical core that remains beyond reach with current computational and numerical capabilities \cite{kim2003overview}. As such, the parametrization of orographic gravity-wave drag and, in particular, answering the question of how good such parametrization schemes are, are ongoing research topics \cite{vosper2020can, van2020constraining}.

Existing methods of parametrizing unresolved orographic gravity-wave drag in NWP models are based on, for example, the study by \citeA{palmer1986alleviation, mcfarlane1987effect, baines1990rationale}, and the scheme introduced by \citeA{lott1997new}. Advances have also been made in characterizing unresolved gravity waves with a raytracing approach, with the Gravity-wave Regional or Global Ray Tracer \cite<GROGRAT;>{marks1995three} and the Multi-Scale Gravity Wave Model \cite<MS-GWaM;>{boloni2021toward, kim2021toward} being two notable examples of gravity-wave raytracers. The latter is coupled to the Icosahedral Non-hydrostatic NWP model \cite<ICON;>{prill2022icon} and supports lateral propagation of non-orographic gravity waves \cite{voelker2023ms}.

In implementing parametrization schemes for unresolved gravity-wave drag, a first problem is presented by the representation of subgrid-scale orography (SSO), i.e., orographic features with a resolution smaller than the effective grid size. While global topography datasets are readily available, deriving a meaningful representation of the SSO for the parametrization scheme is not trivial. Some examples of global topography datasets are the ETOPO1 Global Relief Model \cite{amante2009etopo1}, the Global Multi-Resolution Terrain Elevation Data 2010 \cite<GMTED;>{danielson2011global}, and the Multi-Error-Removed Improved-Terrain Digital Elevation Model \cite<MERIT;>{yamazaki2017merit}.

Having obtained the topographical data from one of the datasets mentioned above, the \citeA{lott1997new} scheme (LM97) would represent the SSO ``via an array of uniformly distributed, elliptical mountains'' with a single length-scale for each grid box \cite{elvidge2019uncertainty}, while \citeA{bacmeister1993mountain} and \citeA{bacmeister1994algorithm} introduced a preprocessing of the topography dataset via a ridge identification method. Naturally, such a coarse approximation is far from physical real-world orography, and more recent attempts at improving SSO representations are, for example, the GROGRAT-based Mountain Wave Model (MWM) introduced by \citeA{rhode2023mountain} for the ridge identification method. However, the SSO representation does not change for both of these methods when the size of the grid cell is varied. 

To circumvent the lack of ``scale awareness'' in SSO representations, \citeA{van2021towards} introduced a spectral representation of SSO that scales more consistently with varying grid sizes. Our work builds upon the ``scale-aware'' innovations by \citeA{van2021towards} of representing SSO in spectral space. Specifically, we introduce a method to approximate the spectrum of SSO with the following major features:
\begin{enumerate}
    \item The scale-aware method is independent of the size of the underlying grid box.
    \item The method works for gridded non-equidistant topographic data distributed in non-quadrilateral grid cells.
    \item The spectral approximation of the SSO requires a hundred or fewer modes to represent orographic features larger than a 5\,km scale.
\end{enumerate}
The two latter features result from the robustness of our spectral approximation method to underlying constraints, e.g., the setup of the dynamical core or the requirements of the gravity-wave parametrization scheme. To highlight this robustness to external constraints, we will refer to the method as the \textit{{constrained spectral approximation method}} (CSAM) from hereon.

We also note that the last feature is particularly relevant to implementing a state-of-the-art gravity-wave raytracing scheme. Raytracing is computationally expensive, and the number of ray-volumes that could be traced, i.e., spectral modes representing the gravity waves at launch height, is limited. Therefore, the compression of a large dimension of physical data points to a limited number of spectral modes is a desirable feature in this pursuit. A brief illustration of the problem is as follows. The GMTED 2010 30-arc\,sec topography dataset comprises approximately $\mathcal{O}(7 \times 10^4)$ physical data points for an icosahedral ICON grid cell of length and height $\sim(160 \times 160)$\,km over the Alaskan Rocky Mountain region. Direct Fourier transform would yield a computationally intractable number of ray-volumes.

To further establish the necessity of the CSAM for SSO, with particular reference to raytracing on a geodesic grid, we elaborate on the three features mentioned above in Section~\ref{sec:motivation}. The detailed implementation of the CSAM is given in Section~\ref{sec:implementation}. Section~\ref{sec:results} provides the experimental setup and documents the results of the method in idealized settings and regional runs on grids similar to the ones used in ICON. The section also elaborates on the principal behavior and capabilities of the method, its potential biases, and limitations. The implementation choices are justified throughout Sections~\ref{sec:implementation} and~\ref{sec:results}. Finally, the limitations of the CSAM and potential avenues for improvement are discussed in Section~\ref{sec:discussions}, and a summary of this work can be found in Section~\ref{sec:conclusions}.

\section{Motivation}
\label{sec:motivation}
Existing methods of SSO representation are based mainly on coarse preprocessing of topographic data. As a result, substantial physical information is lost in setting up a suitable input to the orographic gravity-wave parametrization scheme and contributes to a significant source of error \cite{elvidge2019uncertainty}. Overcoming these limitations inspired the development of the CSAM. In this section, we provide justifications for a spectral representation of the SSO and arguments for the importance of robustness to the external constraints mentioned in the preceding section.

\subsection{Scale-awareness of the subgrid-scale orography}
The effect of orographic gravity waves on atmospheric circulation is a process that is partially resolved and partially parameterized \cite{hindley202018}. The limitation of existing methods to represent SSO robustly across varying grid scales has the most severe implication for coarser grid resolutions used in climate and seasonal projections \cite{van2021towards}, and the limits of existing SSO representations were investigated in \citeA{vosper2015mountain}, \citeA{van2020constraining}, and \citeA{vosper2020can}. One could tune the LM97 and MWM schemes for varying grids, but this would require substantial effort to find the optimal tuning parameters for each grid configuration and size, and the resulting orographic gravity wave drag is susceptible to the assumptions, skill, and experience of the practitioner. Moreover, such an approach would be incongruent with the seamless modeling philosophy that is increasingly adopted by NWP centers. Seamless dynamical cores are an active research topic \cite{wood2014inherently,melvin2019mixed, voitus2019, qaddouri2021implementation}, and they demonstrate potential towards a physical coupling between the dynamical core and data assimilation engine \cite{benacchio2019semi, chew2022one}. A scale-aware approach would circumvent the issues above entirely, and an argument for a spectral-based scale-aware SSO representation similar to the one made here can be found in \citeA{van2021towards}.

A spectral SSO representation is a natural choice for a scale-aware approach, as it encodes information about the full SSO spectrum and does not make any assumptions on the underlying orography. The use of the orographic spectrum to improve the simulation of orographic gravity wave drag was studied by \citeA{garner2005topographic} and \citeA{smith2018gravity}. The former applied spherical harmonics to the approximation of LM97-like SSO, and the latter applied the Fourier transform to better represent the nonlinear effects of flow over orography. Recently, \citeA{van2021towards} extended the use of spectral SSO representation directly in an orographic gravity-wave parametrization that is coupled to the Met Office's United Model. They showed, via simulation runs over the Alaskan Rocky Mountains, that their scale-aware scheme yields an almost constant sum of resolved and parameterized GW momentum fluxes regardless of the grid size. Furthermore, they reported improvement in the excessive wind bias within the polar vortex in the upper troposphere to the lower stratosphere. While the importance of scale awareness in SSO representation is well-established, a significant limitation is the reliance on fast Fourier transform (FFT) that requires underlying quadrilateral grid cells with equidistant topographic data points. 

\subsection{Robustness on unstructured geodesic grids}
There are arguments for the use of geodesic grids in NWP models, with circumventing the pole problem being one of the important advantages \cite{randall2002climate}, and the ICON's triangular icosahedral grid is a notable example of an operational NWP model that uses a geodesic grid. On the other hand, the geodesic dual pentagonal-hexagonal grid has favorable properties such as the conservation of energy and mass, and this grid has been implemented in a variant of the ICON model \cite<ICON-IAP;>{gassmann2011inspection, gassmann2013global}. A summary of the strengths and weaknesses of quadrilateral and non-quadrilateral grids may be found in \citeA{staniforth2012horizontal}. Moreover, to provide a further justification for the development of the CSAM, the response of the dynamical core's underlying grid structure to the effect of unresolved subgrid-scale gravity waves was mentioned by \citeA{staniforth2012horizontal} as a potential topic for further investigation.

For dynamical cores employing a non-quadrilateral geodesic grid, a spectral SSO representation that relies on FFT will not work. A workaround would be to perform an FFT of the SSO over the quadrilateral domain encompassing the geodesic grid cell, but such overlaps in the SSO representation should be avoided \cite{van2021towards}. Variants of the FFT, such as the discrete triangle transform by \citeA{puschel2004discrete}, require an equispaced triangular grid and would not work for geodesic grids on a sphere either. The straightforward CSAM we present in this manuscript provides the approximated SSO spectrum without making assumptions about the underlying grid structure or topographic dataset. The CSAM, therefore, should work for all geodesic grid configurations and resolutions. However, the studies in Section~\ref{sec:results} will only involve triangular grid cells similar to the ones used in the operational ICON NWP model. 

\subsection{Input to a gravity-wave raytracer}
Recent developments in gravity-wave raytracers provide a state-of-the-art weakly nonlinear parametrization of unresolved gravity waves with lateral propagation at a fraction of the computational cost of running fully wave-resolving simulations \cite{voelker2023ms}. However, the implementation of an orographic gravity-wave source in MS-GWaM is impeded by two challenges: (a) An underlying geodesic grid, as mentioned in the previous subsection, and (b) the crucial computational requirement to represent the SSO spectrum with as few spectral modes as possible. We elaborate on point (b) in this subsection.

In a two-dimensional vertical slice, the weakly nonlinear dynamics of gravity waves obtained via a WKB ansatz as a solution to the linearized non-rotating Boussinesq equations have the polarization relations as follows [cf. Eq.~(10) of \citeA{muraschko2015application}],
\begin{equation}
    ( \hat{u}_0, \hat{w}_0, \hat{b}_0/N, \hat{p}_0 ) = a \left( -i \frac{\hat{\omega}}{k}, i \frac{\hat{\omega}}{m}, \frac{N}{m}, -i \frac{\hat{\omega}^2}{k^2} \right),
    \label{eqn:ansatz_soln}
\end{equation}
where $\hat{\omega}$ is the intrinsic frequency; $k$ and $m$ the zonal and vertical wavenumbers respectively; and $( \hat{u}_0, \hat{w}_0, \hat{b}_0, \hat{p}_0 )$ are the leading-order wave solutions of the horizontal velocity, vertical velocity, buoyancy, and pressure of a gravity-wave packet. The complex wave amplitude is denoted by $a$, and $N$ is the Brünt-Väisälä frequency. More details on the weakly nonlinear description of gravity waves can be found in \citeA{achatz2017interaction,achatz2022atmospheric,achatz2023multi}, and the subsequent results can easily be generalized to compressible flow.

From linear theory \cite<see, e.g.,>{nappo2013introduction}, the vertical wave solution at the bottom boundary is given by
\begin{equation}
    \hat{w}_0(z=0) = -i \hat{\omega} \hat{h},
    \label{eqn:lt_bottom_boundary}
\end{equation}
where $z$ denotes the vertical height and $\hat{h}$ the spectral amplitude of the orography. From the definition of the wave-action density, which encodes information on the physical energy of a gravity-wave packet, we have
\begin{equation}
    \mathcal{A} = \frac{E}{\hat{\omega}} = \frac{|\hat{b}_0|^2}{2 \hat{\omega} N^2}.
    \label{eqn:wave_action_density_energy}
\end{equation}
Inserting \eqref{eqn:ansatz_soln} into \eqref{eqn:wave_action_density_energy} and eliminating the amplitude $a$ with \eqref{eqn:lt_bottom_boundary}, we obtain,
\begin{equation}
    \mathcal{A} = - \frac{N^2}{2 \hat{\omega}} |\hat{h}|^2,
    \label{eqn:wave_action_density_final}
\end{equation}
where we note that the power of the SSO spectrum determines the energy of the orographic gravity-wave field at the launch level.

The number of ray-volumes that have to be traced in each grid cell is determined at the launch level by the number of spectral modes required to provide a sufficiently good representation of the SSO spectrum. As such, the fewer the number of spectral modes involved in such a representation, the fewer the number of discrete wave-action densities and ray-volumes needed to be traced in the parametrization process. As raytracing is inherently computationally expensive, the fewer the ray-volumes involved in the process, the better. Therefore, Feature~3 in Section~\ref{sec:intro} is desirable and overcomes the first hurdles of implementing an orographic gravity-wave source in the MS-GWaM raytracer that is necessary even when coupled to a dynamical core with quadrilateral grid cells.

\section{The Constrained Spectral Approximation Method}
\label{sec:implementation}
The CSAM is introduced in this section, and it is based on judiciously interleaving a variant of the least-squares spectral analysis \cite{vanivcek1969approximate, lomb1976least, scargle1982studies} with physically motivated choices and constraints. The CSAM works by progressively compressing the dimension of the spectral space required to represent a large physical dataset while preserving essential physical properties, such as the total power of the spectrum. The progressive compression of the spectral space also allows for encoding additional physical information unavailable to the topographic data in a non-quadrilateral grid cell.

\subsection{Least-squares Fourier fitting}
\label{subsec:lsff}
To avoid confusion with other variants of least-squares spectral analysis, e.g., the Lomb-Scargle Periodogram commonly used in time series analysis \cite{vanderplas2015periodograms}, we refer to the variant used in this work as the ``least-squares Fourier fitting'' (LSFF). The LSFF procedure is documented in this subsection.

Writing out the truncated discrete Fourier representation of a physical orography $h(x,y)$, where $(x,y)$ are the horizontal coordinates, we have
\begin{eqnarray}
h(x,y) &=& 
{\sum_{m=0}^{\mathcal{M}/2-1} \left[ 2 \, \hat{h}_{0,m}^{(r)} \, \cos(l_m y) - 2 \, \hat{h}_{0,m}^{(i)} \, \sin(l_m y) \right]} \nonumber \\
&&+ {\sum_{n=1}^{\mathcal{N}} \, \sum_{{\substack{{m=-\mathcal{M}/2}
}}}^{\mathcal{M}/2-1}} {\left[ 2 \, \hat{h}_{n,m}^{(r)} \, \cos(k_n x + l_m y) \right.} { \left. - 2 \, \hat{h}_{n,m}^{(i)} \,\sin(k_n x + l_m y) \right]},
\label{eqn:final_fourier}
\end{eqnarray}
where the superscripts $(r)$ and $(i)$ represent the real and imaginary components of the Fourier amplitudes $\hat{h}$, and $(k_n,l_m)=(2\pi\,n/L_x , 2\pi\,m/L_y)$ are the discrete horizontal wavenumbers. Here, $n=0,\dots,\mathcal{N}-1$ and $m=-\mathcal{M}/2+1,\dots,\mathcal{M}/2$, with $\mathcal{N}$ and $\mathcal{M}$ being the choice of the truncation limit, and $(L_x,L_y)$ are the horizontal and vertical extent of the physical orography. A detailed derivation of \eqref{eqn:final_fourier} is in~\ref{apx:final_fourier_deriv}.

From \eqref{eqn:final_fourier}, we may set up an optimization problem of the form
\begin{equation}
J[\mathbf{\hat{h}}_{nm}] = \left\lVert \tilde{\pmb{\mathcal{F}}}[\mathbf{\hat{h}}_{nm}] - \mathbf{h}_{xy} \right\rVert^2 + \lambda \left\lVert \mathbf{\hat{h}}_{nm} \right\rVert^2,
\label{eqn:op_prob}
\end{equation}
where $J[\,\cdot\,]$ represents the cost function, and $\left\lVert \, \cdot \, \right\rVert$ is the choice of a suitable norm. The operator $\tilde{\pmb{\mathcal{F}}}$ represents the approximate inverse discrete Fourier transform, and $\mathbf{\hat{h}}_{nm}$ is a vector comprising the Fourier amplitudes $\hat{h}_{n,m}$. The vector $\mathbf{h}_{xy}$ collects all the topographical data points in a grid cell. The second term on the right is chosen to enforce a Tikhonov regularization, with $\lambda$ as the tuning parameter that penalizes overfitting. The choice of an $L_2$-regularizer is discussed in Section~\ref{sec:discussions}.

The minimization of the cost function in Eq. (6) can be found  via linear regression, yielding the unknowns in $\mathbf{\hat{h}}_{nm}$. To do this, take the gradient of the convex loss function with respect to the unknowns,
\begin{equation}
\tilde{\nabla}_{\mathbf{\hat{h}}_{nm}} J[\mathbf{\hat{h}}_{nm}] = 2\, \tilde{\pmb{\mathcal{F}}}^\intercal \tilde{\pmb{\mathcal{F}}}[\mathbf{\hat{h}}_{nm}] - 2 \, \tilde{\pmb{\mathcal{F}}}^\intercal \mathbf{h}_{xy} + 2 \lambda \mathbf{\hat{h}}_{nm},
\label{eqn:grad_loss}
\end{equation}
where $\tilde{\pmb{\mathcal{F}}}^\intercal$ denotes the discrete Fourier transform operator, i.e., the inverse of $\tilde{\pmb{\mathcal{F}}}$. The local minimum may be obtained by setting \eqref{eqn:grad_loss} to zero. By doing so and rearranging, we obtain 
\begin{equation}
\left\lbrace \tilde{\pmb{\mathcal{F}}}^\intercal \tilde{\pmb{\mathcal{F}}} + \lambda \pmb{I} \right\rbrace [\mathbf{\hat{h}}_{nm}] = \tilde{\pmb{\mathcal{F}}}^\intercal \mathbf{h}_{xy}.
\label{eqn:lin_prob}
\end{equation}
Solution of the linear problem in \eqref{eqn:lin_prob} yields $\mathbf{\hat{h}}_{nm}$, the approximate Fourier amplitudes of the orography in a grid cell. 

\subsection{Introducing the constraints}
\label{subsec:intro_constraints}
On a first glance, after an appropriate choice of basic Fourier modes, the solution of \eqref{eqn:lin_prob} is sufficient to obtain an approximation for $\hat{h}$ that is required for the computation of the wave-action density in \eqref{eqn:wave_action_density_final}. This straightforward one-step approach will be referred to as the pure LSFF, and it poses two severe limitations. 

First, the only degree of freedom that determines how well the method performs is the tuning parameter $\lambda$. Too large a value for $\lambda$ leads to over-penalization, and the linear problem solved in \eqref{eqn:lin_prob} deviates from the actual setup we are interested in. Conversely, too small a value leads to overfitting, and the power of the computed SSO spectrum cannot be guaranteed to be physical. The method becomes highly susceptible to the choice of the $\lambda$-value, and the trade-off of over- or under-penalizing the cost function cannot be avoided. To illustrate this point, numerical results involving idealized tests are presented in Section~\ref{subsec:idealized_tests}.

Second, the spectrum of the SSO, while truncated, is nevertheless represented by a dense $(\mathcal{N} \times \mathcal{M})$ number of modes. Assuming a moderately-sized spectral space of $\mathcal{N}=24$ and $\mathcal{M}=48$, the LSFF would provide 1152\,modes as an input to the raytracer. However, this is at least ten times the number of Fourier modes in each grid cell for raytracing with MS-GWaM to be tractable. A $L_1$-regularizer could circumvent this issue but would make the method even more reliant on tuning $\lambda$. This point is discussed further in Section~\ref{sec:discussions}.

The CSAM is introduced as a method relatively robust to \textit{reasonable choices} of $\mathcal{N}$, $\mathcal{M}$, and $\lambda$. While tuning these parameters may yield some improvement, tuning is generally unnecessary to obtain decent error scores. For this reason, we will refer to these parameters as ``free parameters'' instead of ``tuning parameters.'' More details on the robustness of the CSAM and an elaboration on what some of these ``reasonable choices'' means are provided throughout Section~\ref{sec:results}.

\subsubsection*{Algorithm: Constrained Spectral Approximation Method}
In light of these limitations, we subsequently introduce the CSAM as an improvement to a pure LSFF procedure, alongside an illustration of the dimension of the problem at each substep.

For each non-quadrilateral grid cell in a geodesic grid:
\begin{enumerate}
    \itemsep1em 
    \item \label{pt:alg_1} Load the topographic data in the quadrilateral domain encompassing the non-quadrilateral grid cell. For a 30-arc\,sec topography dataset over the Rocky Mountains, this corresponds to $N^{\text{FA}}_\text{data} \sim 7 \times 10^{4}$ data points.
    \item \label{pt:alg_2} Compute the spectrum of the topographic data from Step~\ref{pt:alg_1}. Either FFT or LSFF could be applied to the data in this step. In applying the FFT, the number of spectral modes obtained equals the number of physical data points, i.e., $N_\text{spec}=N_\text{data}$. In applying the LSFF, the free parameters are $\mathcal{N}$, $\mathcal{M}$, and $\lambda_\text{FA}$, and the representation of the SSO is compressed from $N_\text{data}$ to a dense $N_\text{spec}=(\mathcal{N} \times \mathcal{M})$. We will refer to Steps~\ref{pt:alg_1} and~\ref{pt:alg_2} as the First Approximation (FA).
    \item \label{pt:alg_3} From the spectrum computed in Step~\ref{pt:alg_2}, identify the $N_{\text{modes}}$ modes with the largest amplitudes. With a choice of the free parameter $N_\text{modes} \ll N_\text{spec}$, this step compresses the dense spectral space $N_\text{spec}$ to a sparse collection of $N_\text{modes}$ number of spectral modes.
    \item \label{pt:alg_4} Finally, from the $N_\text{modes}$ identified in Step~\ref{pt:alg_3}, do a weakly regularized LSFF on the topographic data in the non-quadrilateral grid cell, i.e., $N^{\text{SA}}_\text{data} \sim 3.5 \times 10^{4}$ for the example dataset. The free parameter in this step is $\lambda_\text{SA}$. Steps~\ref{pt:alg_3} and~\ref{pt:alg_4} are abbreviated as the Second Approximation (SA).
    \setcounter{cnt_cr}{\value{enumi}}
\end{enumerate}

The choice of FFT in Step~\ref{pt:alg_2} is computationally faster. However, LSFF yields better approximations as the spectral space representing the SSO is progressively compressed over two steps, FA and SA. A short study comparing FFT to LSFF in the FA step can be found in \ref{apx:fft_vs_lsff}. 

Aside from the intermediate compression of the spectral space, Step~2 also constrains the spectral modes made available to the LSFF in the SA. Additional large-scale physical information in the quadrilateral grid cell is encoded into the spectrum approximated by the FA step that is not otherwise represented in the right-hand side term of the LSFF in the SA step (see \eqref{eqn:lin_prob} for more details), adding to the robustness and effectiveness of the CSAM.

\subsection{Potential variations}
\label{subsec:potential_variations}
To reduce the number of modes used to represent the SSO, an alternative to the compression of spectral space in a two-step method like the CSAM is the coarse-graining of the spectrum computed from a pure LSFF procedure for a non-quadrilateral grid cell or a choice of FFT or LSFF for a quadrilateral grid cell. However, such a coarse-graining approach is inherently suboptimal. If an averaging kernel is used, the power of the spectrum will always be an underestimation of the true SSO spectrum, and this underestimation will be more severe for larger kernel windows. If a summation kernel is used, the few sparsely distributed spectral modes remaining at the end of the coarse-graining procedure will have artificially large amplitudes, and the resulting spectrum can no longer be guaranteed to be physical. As such, the CSAM is also relevant for quadrilateral grids if the representation of the SSO by a limited number of spectral modes is a constraint required by the parametrization scheme. Numerical tests supporting these arguments have been conducted, but the results are not shown here.

The Gibbs phenomenon can be circumvented by tapering the edges of the topographic data prior to FA and SA. From idealized tests, slight improvement to the error scores may be obtained by applying tapering to the data before FA and SA. To set up a tapering filter that is robust to any non-quadrilateral grid cell, a Shapiro-like filter \cite{shapiro1970smoothing} is applied recursively to a binary mask corresponding to the topographic data in the grid cell, and the Laplace stencil used is the 9-point Oono-Puri \cite{oono1987computationally}. Note that once the filter has been applied to the binary mask, the mask no longer contains only binary values but comprises real numbers ranging from 0 to 1. At each iteration step, the mask values in the grid cell are reset to 1s so that no orographic information within the grid cell is lost as the edges of the domain of interest are smoothed outwards. Once the desired smoothing has been achieved, e.g., after ten recursive applications of the filter, the filtered mask is applied to the topographic data in the grid cell. Before applying tapering, the domain of the topographic data must be padded with the extent of the filter signal, i.e., the topographic data domain has to be at least the size of the smoothed mask. An elaboration on the implementation and illustrations of the tapered orography in a non-quadrilateral grid cell can be found in \ref{apx:tapering}.

Finally, the linear problem \eqref{eqn:lin_prob} could be solved either by a direct Cholesky factorization or via iterative solvers such as the generalized minimal residual method \cite<GMRES;>{saad1986gmres}. In the CSAM, using the latter with the GMRES solver is preferred (supporting results are not shown), as it is faster than a direct solver with minimal loss of accuracy. Moreover, the issue of the Gibbs phenomenon may be mitigated by setting a sufficiently large tolerance for the iterative solver, e.g., $10^{-5}$. A larger tolerance implicitly avoids fitting high-frequency signals associated with the Gibbs phenomenon in the linear regression. However, using an iterative solver introduces additional free parameters to be determined by the practitioner.

Having tested the potential variations to the CSAM, the results presented in the subsequent sections are generated with the following configuration: No coarse-graining of the spectral space is applied, and tapering over ten grid cells is applied to the boundary of the FA and SA domains for studies with real topographic data. See \ref{apx:tapering} for the implementation details. Finally, a GMRES iterative solver with a tolerance of $10^{-5}$ is used in solving \eqref{eqn:lin_prob}. 


\subsection{Iterative refinement}
\label{subsec:it_refinement}
The CSAM presented so far is an interleaving of the LSFF with physically motivated choices to achieve the major features mentioned above. The principal capability of the method may be further cemented by better characterizing its behavior. Specifically, this subsection introduces an extension to the CSAM such that the approximation could be iteratively improved if a reference solution is known, and the method is shown to behave in line with expectations. We refer to this extension as the iterative refinement (IR).

We assume a scenario where the target of the cost function $J[\mathbf{\hat{h}}_{nm}]$ is known. Note that the target does not necessarily have to be $\mathbf{\hat{h}}_{nm}$, and knowing a derived quantity, such as the pseudo-momentum fluxes, is sufficient. Furthermore, we assume that the linear regression problem is adequately regularized and that any overestimation of the target quantity is not due to catastrophic overfitting.

Now, given the assumptions above, an approximated spectrum $\mathbf{\hat{h}}_{nm}$ that overestimates the target is due to spurious small-scale noise introduced by the approximation procedure. Conversely, if the approximation leads to underestimating the true solution, crucial higher-frequency spectral modes were not adequately captured in the two-step approach. Severe over- and underestimation could be rectified by a refinement step (RF) as follows. Recover from the approximated FA spectrum $\mathbf{\hat{h}}_{nm}$ the physical approximation of the orography in the quadrilateral grid cell, 
\begin{equation}
    \mathbf{h}_{xy}^\text{approx} = \tilde{\pmb{\mathcal{F}}}[\mathbf{\hat{h}}_{nm}].
    \label{eqn:recovery}
\end{equation}
The choice of using the FA spectrum in this step is further discussed in Section~\ref{subsec:iterative_refinement_results}. Take the residual $\mathbf{h}_{xy}^\text{res}$ of the reconstructed orography from the original orography, 
\begin{equation}
    \mathbf{{h}}_{xy}^\text{res} = \mp \left(\mathbf{{h}}_{xy} - \mathbf{{h}}_{xy}^\text{approx}\right),
    \label{eqn:phys_residual}
\end{equation}
where, if the negative sign is chosen, the residual captures the spurious noise contributing to an overestimation. If the positive sign is chosen, the residual captures the small-scale features not yet represented by the approximated spectrum. 

The CSAM algorithm is then applied to the residual, and the spectrum obtained from this refinement step is add to the approximation spectrum. Finally, the effective SSO spectrum is obtained by choosing the $N_\text{modes}$-largest modes from the approximated spectrum and the spectrum obtained from the refinement step. The IR procedure can be repeated until a desired error bound is met.

A summary of the IR algorithm is provided below, and numerical results from applying this extension are in Section~\ref{subsec:iterative_refinement_results}.

\subsubsection*{Algorithm: Iterative refinement}
\begin{enumerate}
    \setcounter{enumi}{\value{cnt_cr}}
    \itemsep1em
    \item \label{pt:alg_5} If the approximated spectrum over- or underestimates the reference beyond a certain tolerance, say $\texttt{tol=}20\%$ relative error, recover the approximated physical FA orography in the quadrilateral domain from Step~\ref{pt:alg_2} and compute the residual in \eqref{eqn:phys_residual}. Otherwise, no iterative refinement is necessary, as the first CSAM approximation is deemed ``good enough.''
    \item \label{pt:alg_6} Repeat the CSAM, i.e., Steps~\ref{pt:alg_1} to~\ref{pt:alg_4}, with $\mathbf{\hat{h}}_{xy}^\text{res}$ as the right-hand side of the linear problem~\eqref{eqn:lin_prob}. The CSAM free parameters may optionally be varied in the iterative refinement procedure.
    \item \label{pt:alg_7} The spectral modes obtained in Steps~\ref{pt:alg_4} and~\ref{pt:alg_6} are added together.
    \item \label{pt:alg_8} Select the $N_\text{modes}$ spectral modes with the largest amplitudes. This yields the effective SSO spectrum after RF.
    \item Repeat Steps~\ref{pt:alg_5} to~\ref{pt:alg_8} until the desired relative error tolerance \texttt{tol} is achieved.
\end{enumerate}

Note that the subsequent residual computed in Step~\ref{pt:alg_5} is the difference between the reference orography and the last approximate FA orography reconstructed from Step~\ref{pt:alg_6}. This ensures that only noise or features of progressively smaller scales are captured.

\section{Results}
\label{sec:results}
This section presents numerical results involving the CSAM, and it is split into six segments. The first segment establishes the effectiveness of the CSAM over a pure LSFF procedure in an idealized setting. The second segment details the experimental setup involving real-world topographic datasets, and the third segment features regional runs over non-quadrilateral grid cells comprising a Delaunay triangulation of the regional domain. Such runs allow for comparison with reference results obtained from FFT, and the effectiveness of the CSAM can be established via a suitable error metric. The fourth to sixth segments seek to better characterize the limitations and behavior of the CSAM.

\subsection{An idealized test case}
\label{subsec:idealized_tests}
\begin{figure}[ht]
    \centering
    \includegraphics[width=\textwidth]{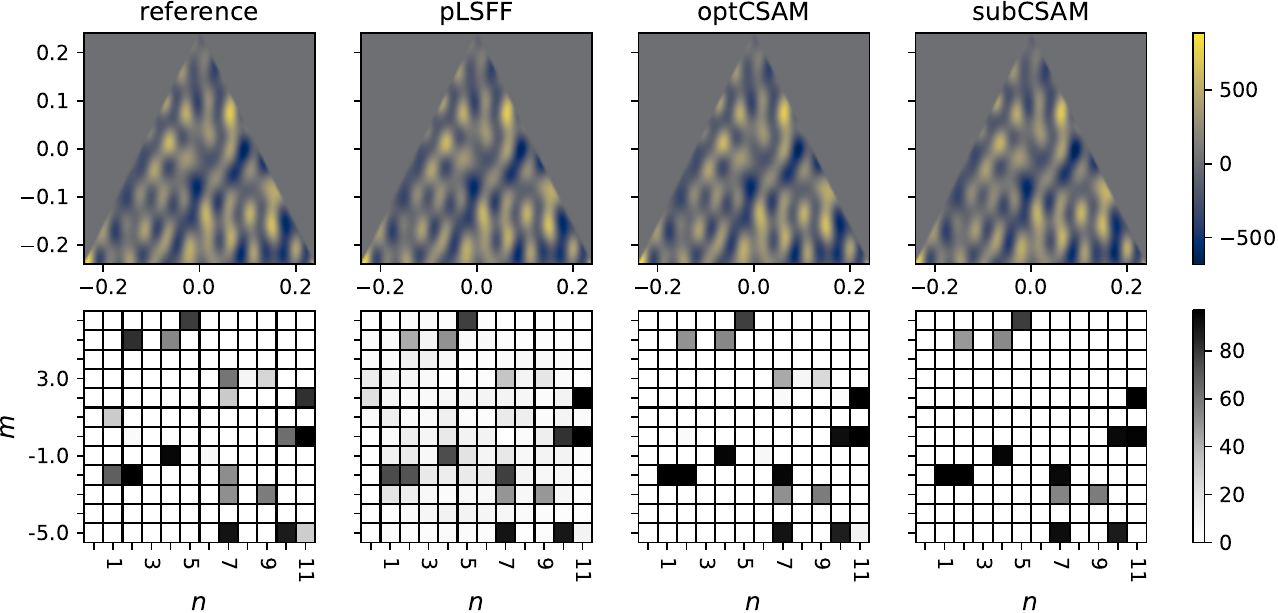}
    \caption{Column wise: Reference solution (first column); results from the pLSFF, optCSAM, and subCSAM experiments (second, third, and fourth columns, respectively). From left to right: The top row depicts the artificially generated reference terrain and the physical reconstruction from the approximated spectrum for each of the three experiments; the bottom row depicts the reference spectrum and the approximated spectra obtained from the three experiments. Note that the notion of physical units is irrelevant in the idealized tests, and the quantities depicted here can be considered dimensionless. Furthermore, note that the pLSFF spectrum (bottom row; second column) is a dense depiction comprising many small but non-zero entries, while the optCSAM and subCSAM spectra (bottom row; third and fourth columns) are sparse depictions.}
    \label{fig:idealized_plots}
\end{figure}

In this section, we investigate the effectiveness of the CSAM against a pure LSFF procedure. The experiments are conducted on a simple isosceles triangle as the grid cell with artificially generated terrain. The artificial terrain is generated as a superposition of randomly generated sinusoidal functions, giving a set of reference spectral solutions,
\begin{equation}
    h^\text{art}(x,y) = \sum_{i=1}^{\mathfrak{S}} \alpha_i \, \mathfrak{A}_{i} \cos \left( \frac{2\pi}{\Delta x}\mathfrak{K}_i x + \frac{2\pi}{\Delta y}\mathfrak{L}_i y \right) + (1-\alpha_i) \, \mathfrak{B}_{i} \sin \left( \frac{2\pi}{\Delta x}\mathfrak{K}_i x + \frac{2\pi}{\Delta y}\mathfrak{L}_i y \right).
    \label{eqn:art_terrain}
\end{equation}
For each $i$, $\mathfrak{K}$ is a randomly generated integer between $[0,12)$, and $\mathfrak{L}$ between $[-5,7)$. This corresponds to $\mathcal{N},\mathcal{M}=12$ in \eqref{eqn:final_fourier}. Any overlapping tuple pair is removed. The amplitudes $\mathfrak{A}$ and $\mathfrak{B}$ are random real numbers between $[0,100)$, and $\alpha$ is either 0 or 1, selected at random. The random number generator is seeded for reproducibility, with the number of active spectral modes $\mathfrak{S}=22$ for the experiment presented here. This randomly generated terrain in the isosceles triangle and its corresponding reference spectrum is depicted in the leftmost column of Figure~\ref{fig:idealized_plots}. 

Now that we have an artificial reference solution, three experiments are conducted and compared against this reference. The first experiment, abbreviated as pLSFF from hereon, is a pure LSFF-only run with $\lambda=8\times10^{-5}$. This penalty term is almost optimal, as doubling or halving its value increases the $L_2$-error in the spectrum. The second experiment is an optimal CSAM run (optCSAM) where the free parameters used in the CSAM algorithm correspond to those used in the artificial generation of \eqref{eqn:art_terrain}, i.e., $\mathcal{N},\mathcal{M}=12$ and $N_\text{modes}=\mathfrak{S}$. Finally, a suboptimal CSAM run (subCSAM) is conducted where fewer number of modes are identified in SA than the true $\mathfrak{S}$, i.e., $N_\text{modes}=14$.

Note that in both the CSAM runs, $\lambda_\text{FA}=0.1$ and $\lambda_\text{SA}=10^{-6}$. No tuning was done on these free parameters except for a judicious choice guided by the following principle: Regularization is generally irrelevant for FA, and a strong regularization could be used as we are only interested in correctly identifying the most relevant modes over a quadrilateral domain, while only a small penalty term is required in SA, as the size of the linear problem has been drastically reduced at this point. 

Referring to Figure~\ref{fig:idealized_plots}, the top row depicting the reconstructed physical orographies obtained from the approximated spectra for all three experiments is qualitatively close to the reference solution. However, the spectra depicted in the bottom row differ significantly. In the pLSFF case (second panel from the left), the approximated spectrum is not sparse, and spectral modes with zero amplitudes in the reference solution have tiny but non-zero amplitudes here. On the other hand, the amplitudes of some modes are underestimated. This underestimation of the spectral amplitudes can also be seen in the optCSAM case (third panel from the left). For example, the amplitude of the $(n,m)=(1,1)$ mode is substantially underestimated. The subCSAM experiment (rightmost panel) is limited by a fewer-than-optimal number of spectral modes, as seen in the greater sparsity of the approximated spectrum compared to the optCSAM case.

\begin{figure}
    \centering
    \includegraphics[scale=0.6]{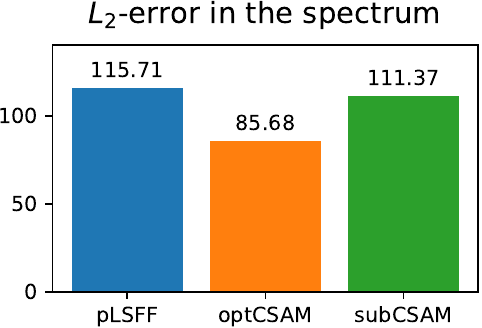} \qquad
    \includegraphics[scale=0.6]{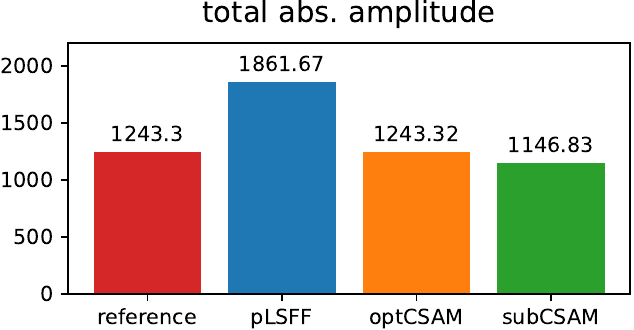}
    \caption{(left panel) $L_2$-error in the experimentally approximated spectra from the reference spectrum. (right panel) The sum of the absolute amplitudes over all spectral modes in the reference and experimentally approximated spectra. As in Figure~\ref{fig:idealized_plots}, the quantities depicted are dimensionless.}
    \label{fig:idealized_errors}
\end{figure}

Figure~\ref{fig:idealized_errors} quantifies the errors in the spectra depicted in Figure~\ref{fig:idealized_plots}. The left panel shows the $L_2$-error in the experimentally obtained spectra from the reference solution, i.e.
\begin{equation}
    L_2\text{-error} = \sqrt{ \sum_{n}^{\mathcal{N}} \sum_{m}^{\mathcal{M}} \left( \mathbf{\hat{h}}_{nm}^{\text{experiment}} - \mathbf{\hat{h}}_{nm}^{\text{reference}} \right)^2 }.
\end{equation}
Even with an optimally tuned penalty term $\lambda$, the pLSFF error is slightly larger than the suboptimal subCSAM run that reproduces the spectrum with just 14 spectral modes. This higher error in the pLSFF run is likely due to the collection of small spurious modes as mentioned above. The optCSAM experiment produces the smallest error, as should be expected when inspecting the experimental spectra in Figure~\ref{fig:idealized_plots}. While the $L_2$-error tells us how close the spectra are, the total absolute amplitude of the spectrum is ultimately the quantity of interest in the spectral representation of SSO, and it is defined as
\begin{eqnarray}
    \text{total abs. amplitude} = \sum_{n}^{\mathcal{N}} \sum_{m}^{\mathcal{M}} \left\lvert \mathbf{\hat{h}}_{nm} \right\rvert
    \label{eqn:total_abs_ampl}
\end{eqnarray}
Recall from \eqref{eqn:wave_action_density_final} that the total absolute amplitude in \eqref{eqn:total_abs_ampl} determines the energy of the gravity-wave packet at the launch level, and this quantity must be as close as possible to the physical energy.

The right panel of Figure~\ref{fig:idealized_errors} shows most glaringly the limitations of the pLSFF run. The contribution of the spurious spectral modes leads to an approximately 50\% overestimation of the reference total amplitude, and parametrization schemes employing such an SSO representation will unavoidably lead to overly energetic orographic gravity waves. The optCSAM run can reproduce the total reference amplitude up to 0.01\% error. This almost exact reproduction is not surprising, as we have set up the CSAM with the optimal free parameters that we happen to know a priori. Finally, with just 64\% of the reference number of spectral modes, the subCSAM run reproduces a spectrum with an 8\% underestimation of the true total amplitude. Such an underestimation is unavoidable as the CSAM attempts to reproduce the true spectrum with fewer degrees of freedom. Nevertheless, the method is robust to such limitations, and the resultant error is small. 

\begin{figure}
    \centering
    \includegraphics[width=\textwidth]{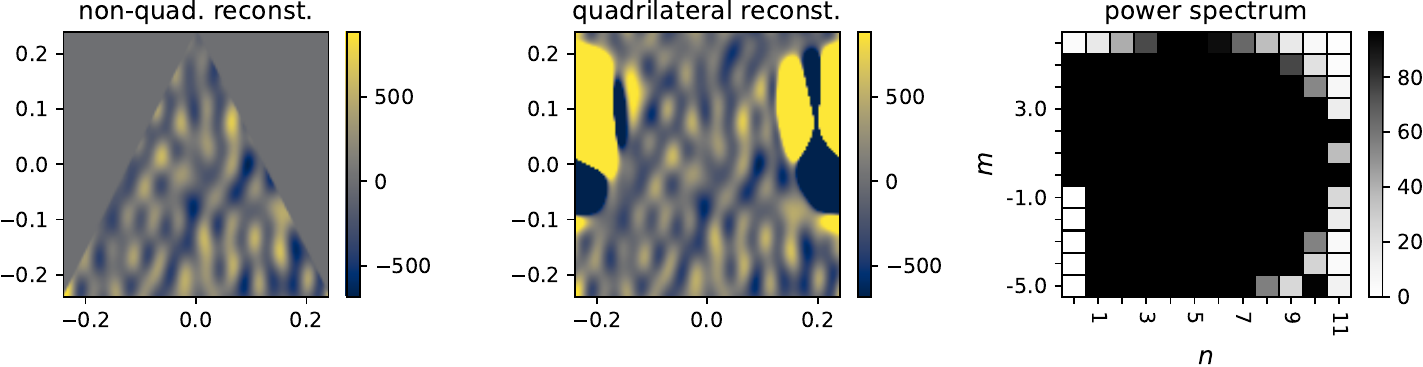}
    \caption{Unregularized pLSFF physical reconstruction in the isosceles triangle (left), over the encompassing quadrilateral domain (middle), and the corresponding power spectrum (right). The extent of the color bars are identical to Figure~\ref{fig:idealized_plots} for comparison. Catastrophic overfitting occurs without regularization.}
    \label{fig:overfitting_issue}
\end{figure}

A pLSFF experiment with no regularization, i.e., $\lambda=0$, was conducted alongside the experiments above. Regularization is vital as otherwise catastrophic overfitting will occur. Referring to Figure~\ref{fig:overfitting_issue}, the spectrum (rightmost panel) reproduces a faithful physical reconstruction inside the domain of interest, the isosceles triangle (leftmost panel). However, the spectrum does so by the superposition of many large amplitude modes. This overfitting results in an unphysical reconstruction outside the domain of interest, e.g., within the encompassing quadrilateral domain (middle panel). As a result, the $L_2$-error for this run without regularization is 163745, i.e., about three orders of magnitude larger than the regularized pLSFF run. The total absolute amplitude is 1107786, corresponding to an 89001\% overestimation of the reference value. The pure LSFF procedure fails if no regularization is applied to the fitting of data distributed over a non-quadrilateral grid cell. 

These experiments demonstrate that, under ideal conditions where the grid cell is relatively simple, and the underlying terrain is a superposition of sinusoidal waves, the CSAM outperforms a pure regularized LSFF procedure that has been almost optimally tuned. Furthermore, even when the free parameters chosen deviate from the optimal reference values, CSAM outperforms the pure LSFF experiment. Having established the capabilities of the CSAM under an idealized setting, we apply it to more challenging real-world orography as a next step. 

\subsection{Experimental setup}
The experimental setup used in conducting simulation runs with real-world topography is detailed in this section. 

\subsubsection{The digital elevation model and idealized domain decomposition}
The topographic dataset used in this work is the high resolution MERIT digital elevation model (DEM) \cite{yamazaki2017merit} at 3~arc\,sec resolution. However, a coarse-graining pre-processing step is applied to obtain an effective 30-arc\,sec dataset. The coarse-graining applied here is akin to applying a moving average window without overlaps. This lower resolution topographic data does not significantly affect the computed results, but it decreases the computational needs substantially by reducing the dimensions of the linear problem in \eqref{eqn:lin_prob}. This dimensional reduction has the beneficial side effect that the linear regression problem is more robust to potential overfitting.

\begin{figure}[t]
    \centering
    \includegraphics[width=\textwidth]{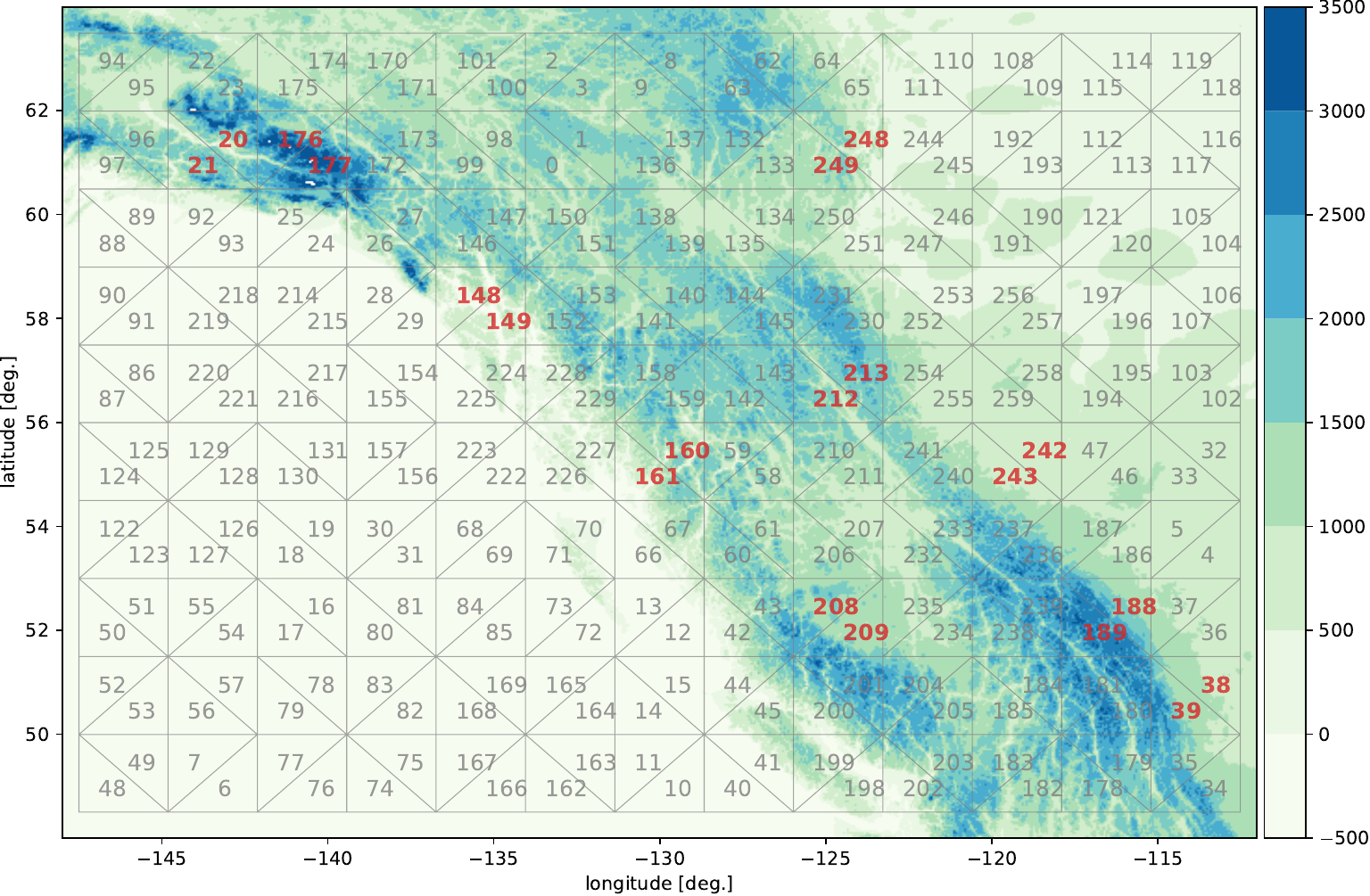}
    \caption{Delaunay decomposition of the Alaskan Rocky Mountains region with an extent of W\,$112^\circ$ to W\,$148^\circ$ and N\,$48^\circ$ to N\,$64^\circ$. Each quadrilateral grid cell comprising a triangle pair is approximately $(160 \times 160)$\,km. The triangle pairs with indices highlighted in red belong to a randomly selected subset of grid cells that are studied in \ref{apx:fft_vs_lsff}. The color bar denotes the orographic elevation in meters [m].}
    \label{fig:delaunay}
\end{figure}

The regional runs depicted in this section are decomposed into triangular grid cells via a Delaunay decomposition. This idealized domain decomposition is such that two triangular grid cells correspond to a quadrilateral grid, and the results computed over the non-quadrilateral triangular grid pair can be compared to a reference value computed via FFT over the corresponding quadrilateral grid cell. Therefore, the CSAM's effectiveness can be quantified with this setup. In the Delaunay decomposition, the grid size is chosen such that each quadrilateral grid cell corresponding to a triangle pair is approximately ($160 \times 160$)\,km, which is comparable to the ICON R2B4 grid \cite{prill2022icon}, or approximately ($80 \times 80$)\,km which is comparable to the R2B5 grid. Refer to Figure~\ref{fig:delaunay} for an example of an idealized domain decomposition over the Alaskan Rocky Mountains.

The CSAM is only applied to grid cells with non-trivial orography; e.g., grid cells predominantly over the ocean are ignored. Specifically, the CSAM is only applied to grid cells where more than 5\% of its topographic data points lie 50\,cm above sea level. Grid cells that meet this requirement for the CSAM are referred to as land cells. Furthermore, elevations below -500\,m are clipped to -500\,m to remove the effect of non-land terrain, e.g., seabeds, that may be present in the topographic dataset. Following \citeA[cf. the Appendix]{van2021towards}, the orography in each grid cell is deplaned by subtracting its mean, and a 5\,km smoother is applied to the topographic data such that features smaller than this length scale are removed. The latter is to remove spectral contributions from small-scale features that do not excite internal orographic gravity waves. See \citeA{williams2020addressing} for more details. 

Finally, in each grid cell, the latitude-longitude coordinates of the topographic data points are converted to planar $(x,y)$-distances from the origin in meters via a Plate Carrée projection. The origin is always considered the southernmost and westernmost topographic data point. The topographic data points are then interpolated onto a regular equidistant grid whenever a quadrilateral domain is available, i.e., in the FFT reference value computation and the FA step of the algorithm. Note that while interpolation is done in the FA step, it is not strictly necessary. No interpolation is done for the SA step on a non-quadrilateral domain, and the data points used in the right-hand side of \eqref{eqn:lin_prob} are structured but non-equidistant. While interpolation inside the non-quadrilateral grid may improve the CSAM's performance, we consciously decided on the more challenging scenario of working with non-equidistant topographic data. Further elaborations are discussed in Section~\ref{sec:discussions}.

\subsubsection{Idealized pseudo-momentum fluxes}
Following \citeA[cf. Eq. (35)]{muraschko2015application}, we compute the horizontal mean of the horizontal pseudo-momentum flux over the discrete spectral modes in the grid cell as follows,
\begin{equation}
    \mathcal{P}:= \overline{u^\prime w^\prime} = \sum_{j} \mathcal{A}_j k_j c_{gz,j}, 
    \label{eqn:horizontal_pmf}
\end{equation}
where $j$ indexes the spectral modes in a grid cell, and $\mathcal{A}$ is the wave-action density obtained from \eqref{eqn:wave_action_density_final} which is dependent on the approximated spectrum of the SSO. The vertical group velocity of the spectral mode $c_{gz}$ is computed from
\begin{equation}
    c_{gz} = \frac{N (k^2+l^2)^{1/2} m}{(k^2 + l^2 + m^2)^{3/2}},
\end{equation}
where the positive branch of the vertical group velocity is chosen. This assumption is valid, as the orographic gravity waves must travel upwards from the launch level. The vertical wavenumber $m$ may be computed from the dispersion relation
\begin{equation}
    \hat{\omega}^2 = (\omega - kU - lV)^2 = \frac{N^2 (k^2+l^2)}{k^2+ l^2+ m^2},
    \label{eqn:dispersion_relation}
\end{equation}
where $\omega$ is the apparent frequency and is zero in the treatment of orographic gravity waves, $U$ the zonal background wind, and $V$ the meridional background wind. Note that the discrete wavenumber, e.g., $k_n$, has the form
\begin{equation}
    k_n = \frac{2\pi n }{\Delta x N_x}, \qquad n=0,\dots,N \,,
\end{equation}
where in the absence of an equidistant distribution of topographic data points in the grid cell, the longitudinal spacing between data points $\Delta x$ is taken as the average spacing over all data points in the grid cell. The total number of data points in the longitudinal direction, $N_x$, is taken over the largest longitudinal extent in a non-quadrilateral grid cell. Similar assumptions for the latitudinal direction are made in the computation of $l_m$.

The pseudo-momentum flux (PMF) computed in \eqref{eqn:horizontal_pmf} is the quantity that will be used to gauge the performance of the CSAM for tests involving real-world topographic data. As the background winds $(U,V)$ and the Brunt-Väisälä frequency $N$ are prescribed in the computation of \eqref{eqn:dispersion_relation}, and the effect of the Coriolis forces are neglected, the computed pseudo-momentum fluxes are to be taken as idealized values.

\subsubsection{The error metrics}
\label{subsec:error_metrics}
To evaluate the performance of the CSAM via PMFs, two error metrics are introduced below. First, the idealized PMF is computed for each triangle in a triangle pair, and the total effective PMF contribution  $\mathcal{P}^{\text{eff}}$ in a quadrilateral grid cell is taken as the sum of the idealized PMF pair, i.e.,
\begin{equation}
     \mathcal{P}^{\text{eff}} =  \mathcal{P}^{\text{T1}} + \mathcal{P}^{\text{T2}},
     \label{eqn:pmf_eff}
\end{equation}
where T1 and T2 denote the respective triangle in the triangle pair. \ref{apx:pmf_computation} provides a justification for this computation of the effective pseudo-momentum flux. The approximated $\mathcal{P}^{\text{eff}}$ is compared to the reference PMF $\mathcal{P}^{\text{ref}}$ obtained from the FFT computation. The local relative error (LRE) is computed as follows,
\begin{equation}
    \text{LRE} = \mathcal{P}^{\text{eff}} / \mathcal{P}^{\text{ref}} - 1.0,
\end{equation}
and the maximum relative error (MRE) is computed as,
\begin{equation}
    \text{MRE} = \frac{\mathcal{P}^{\text{eff}} - \mathcal{P}^{\text{ref}}}{\mathcal{P}^{\text{max}}},
\end{equation}
where $\mathcal{P}^{\text{max}}$ is the maximum PMF in the regional domain.

The LRE metric will be used to assess how well the CSAM performs. However, the LRE may be considerable even when the PMF contribution of a grid cell is small compared to its neighbors. Thus, in the subsequent regional study, the MRE is used to determine locations in the regional domain where the CSAM struggles to reproduce the most physically relevant PMFs.

The error scaling may be observed from rewriting \eqref{eqn:horizontal_pmf} in terms of the intrinsic frequency $\hat{\omega}$, which is dependent on the background wind $(U,V)$. Doing so, one obtains $\mathcal{P} \sim \hat{\omega}^2 |\hat{h}|^2$. As such, any error in the approximated spectral amplitudes will scale quadratically with the background wind. Sufficiently good SSO representation is therefore necessary to keep the error small, especially in the presence of strong background winds. The performance of the CSAM under a strong background wind condition is investigated in Section~\ref{subsec:sensitivity}.

\subsection{Regional runs over the Alaskan Rocky Mountains}
Following \citeA{van2021towards}, regional runs over the Alaskan Rocky Mountains, as depicted in Figure~\ref{fig:delaunay}, are studied in this section. The experimental parameters used are: $\lambda_{\text{FA}}=0.1$; $\lambda_{\text{SA}}=0.1$; $U=10$\,m\,s$^{-1}$ $V=0$\,m\,s$^{-1}$; and a tapering over ten grid cells is used in both FA and SA; see \ref{apx:tapering} for more details. Note the significantly larger $\lambda_{\text{SA}}$ value used here as compared to the idealized experiments in Section~\ref{subsec:idealized_tests}. This choice is motivated by the more complex, non-equidistant underlying real-world topographic data comprising more spectral modes that may not be adequately captured by the discrete truncated spectra involved in the least-square spectral analysis steps. The extent of the compressed spectral domain in the FA step is determined by $\mathcal{N}=\text{(domain length)}/ 5\,\text{km}$ and $\mathcal{M}=2\mathcal{N}$ to ensure $(k,l)$-symmetry in the spectral domain.

\subsubsection{Coarse Delaunay triangulation (approximately R2B4)}
\label{subsec:coarse_study}
The coarser ICON R2B4 grid is such that each non-quadrilateral grid cell fits into an approximately $(160 \times 160)$\,km quadrilateral grid and is suitable for climate simulations. The Delaunay triangulation produces non-quadrilateral grid cells of similar sizes in this section. Here, the choice of $N_{\text{modes}}=100$ is used with optimal $(\mathcal{N},\mathcal{M}) = (32,64)$. As filtering has removed features smaller than 5\,km, the optimal number of dense spectral modes for the FA step may be computed as $\mathcal{N}=160\text{\,km} / 5\text{\,km}=32$.

\begin{figure}[ht]
    \centering
    \includegraphics[width=1.0\textwidth]{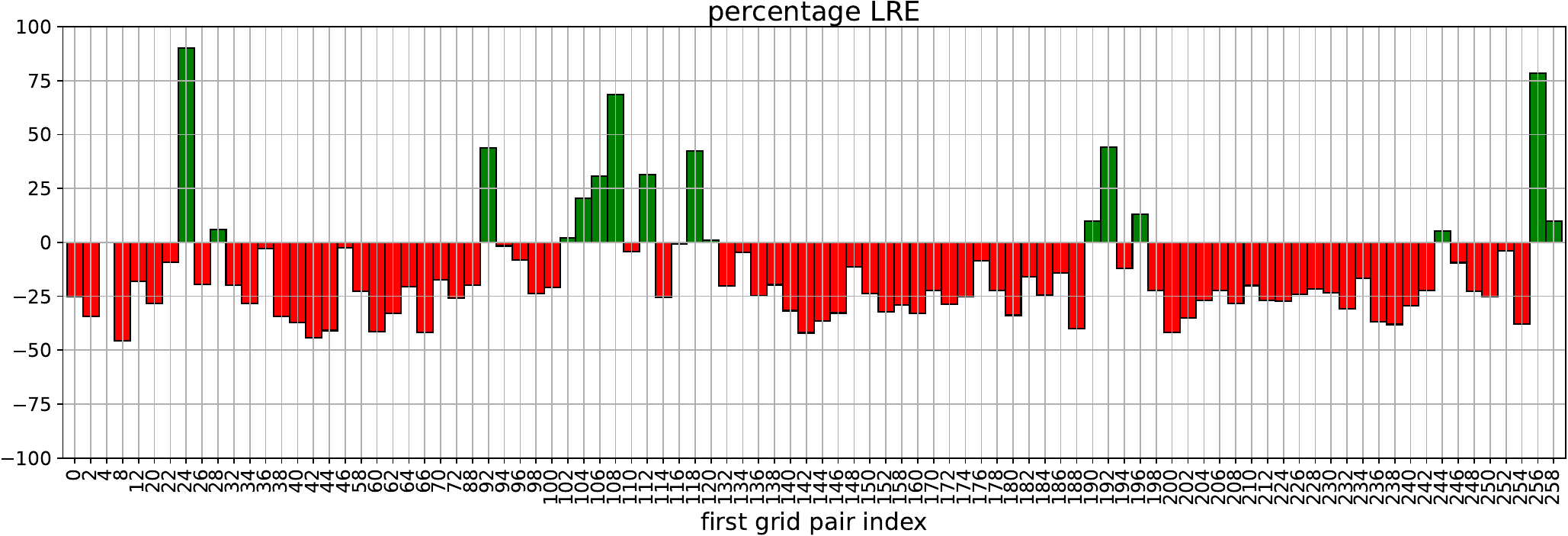}
    \caption{Percentage local relative error (LRE) for each land grid-cell pair in the coarse triangulation of the Alaskan Rocky Mountain region. Overestimation of the reference idealized PMF by the CSAM has a positive error (green bars) and underestimation a negative error (red bars). Note that the indices on the $x$-axis label only the first index of the triangle pair. The average absolute LRE is $25.30\%$.}
    \label{fig:percent_lre}
\end{figure}

The percentage LRE for each land triangle pair is depicted in Figure~\ref{fig:percent_lre} with an average absolute LRE of $25.30\%$. Unsurprisingly, the approximated idealized PMF underestimates most grid pairs, as the CSAM struggles to encode as much topographic information as possible with just 100~effective spectral modes. The idealized PMF is overestimated in a few grid cells, but recall that the LRE is a harsh error metric, as small differences in the idealized PMFs can contribute to large errors. Therefore, it remains to be determined if such overestimation is a valid cause for concern.

\begin{figure}[ht]
    \centering
    \includegraphics[width=1.0\textwidth]{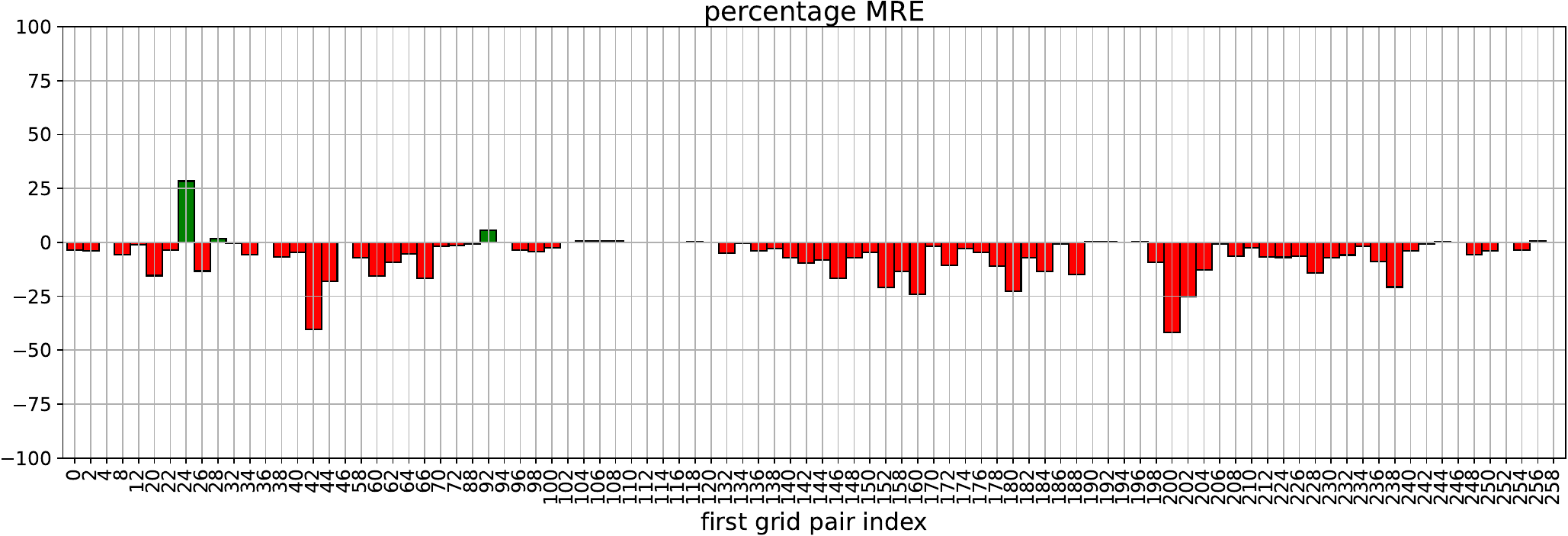}
    \caption{Percentage maximum relative error (MRE) for each land grid-cell pair in the coarse triangulation of the Alaskan Rocky Mountain region. The average absolute MRE is $6.82\%$. See Figure~\ref{fig:percent_lre} for more information on this plot.}
    \label{fig:percent_mre}
\end{figure}

Figure~\ref{fig:percent_mre} depicts the MREs for the land cells. The average MRE is $6.82\%$, recalling that the error is computed relative to the largest reference idealized PMF in the regional domain. From a regional perspective, the underestimation of the reference PMF becomes generally less severe, and only two grid-cell pairs have close to a $-50\%$ error. Nevertheless, the grid-cell pair with indices $(24,25)$ still exhibits substantial overestimation of the idealized PMF, and this case will be studied in Section~\ref{subsec:potential_biases}. 

\begin{figure}[ht]
    \centering
    \includegraphics[width=1.0\textwidth]{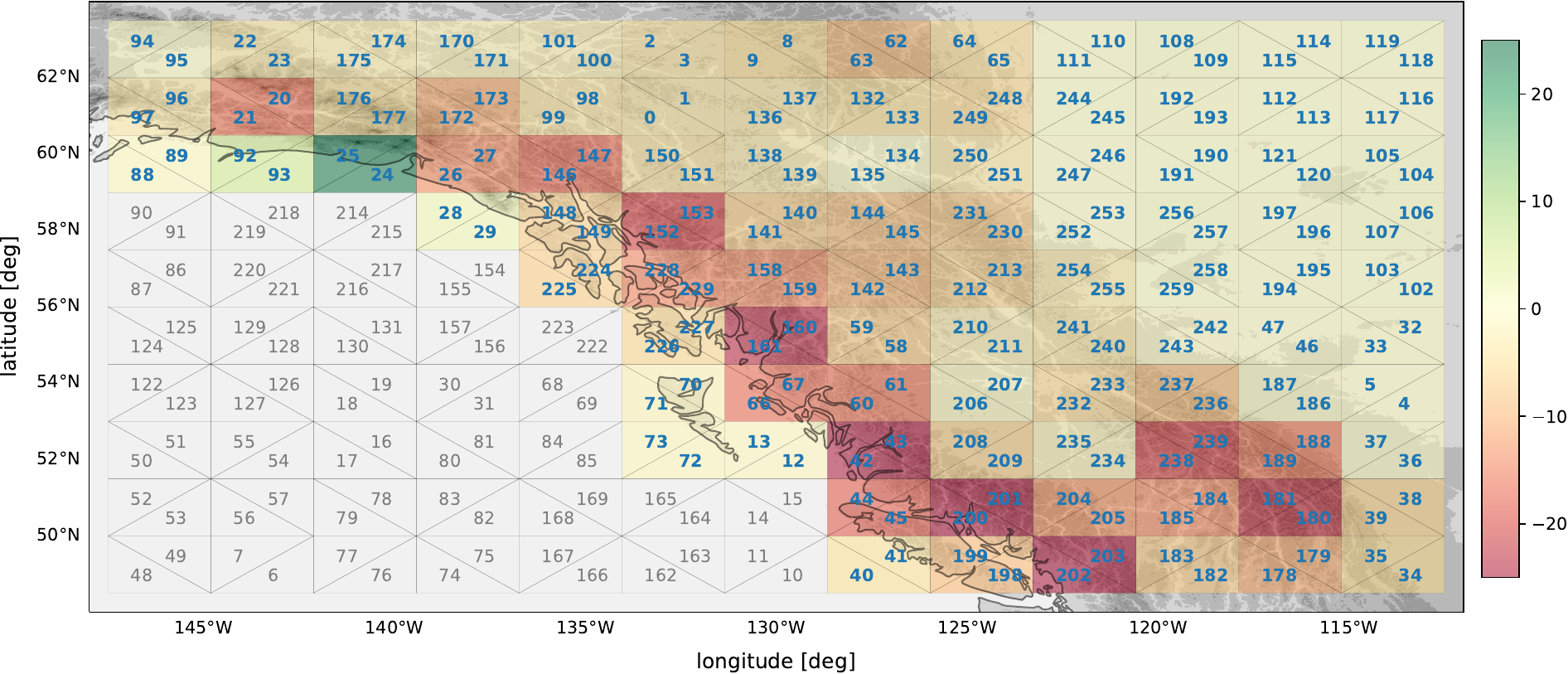}
    \caption{Percentage maximum relative error distribution over the Alaskan Rocky Mountain region for each coarse quadrilateral grid cell. Severe over- and underestimation are most pronounced over sharp gradients and complex topographies; see accompanying text for an elaboration. The color bar range is $[-25\%,25\%]$~error.}
    \label{fig:error_plot}
\end{figure}

Finally, the MREs are plotted on the regional domain in Figure~\ref{fig:error_plot}, and the distribution of the errors becomes clear. Underestimation of the reference idealized PMFs occurs over complex terrain, particularly sharp gradients. E.g., cell pairs (238,239) and (180,181) are over sharp cliffs, and the cluster of severe underestimation around the cell pair (200,201) features mountainous coastal terrain. Again, these observations highlight the limitations of the CSAM. Despite the gradual compression of the spectrum via linear regression, high-frequency features cannot be adequately captured, and the method performs relatively poorly over such terrain features.

\subsubsection{Fine Delaunay triangulation (approximately R2B5)}
A feature of the CSAM is its scale awareness, and the study presented below is for triangle pairs that correspond to an approximately $(80 \times 80)$\,km quadrilateral grid. This is comparable to the ICON R2B5 grid. Here, the choice of $N_{\text{modes}}=50$ is used with $(\mathcal{N},\mathcal{M}) = (16,32)$. The smaller number of effective spectral modes is possible due to the smaller grid cell size. 

For this finer grid size, plots similar to Figures~\ref{fig:percent_lre} and~\ref{fig:percent_mre} are not illustrative due to the numerous grid cells in the triangulation of the domain. However, we note that the average absolute LRE is $20.09\%$, and the average absolute MRE is $2.91$\%. This lower average error may be attributed to the smaller grid size. The CSAM algorithm generally performs better for smaller grid cells, as the topographic information approximated by the spectrum in each grid cell has to represent fewer physically relevant length scales than in larger grid cells.

\begin{figure}[ht]
    \centering
    \includegraphics[width=1.0\textwidth]{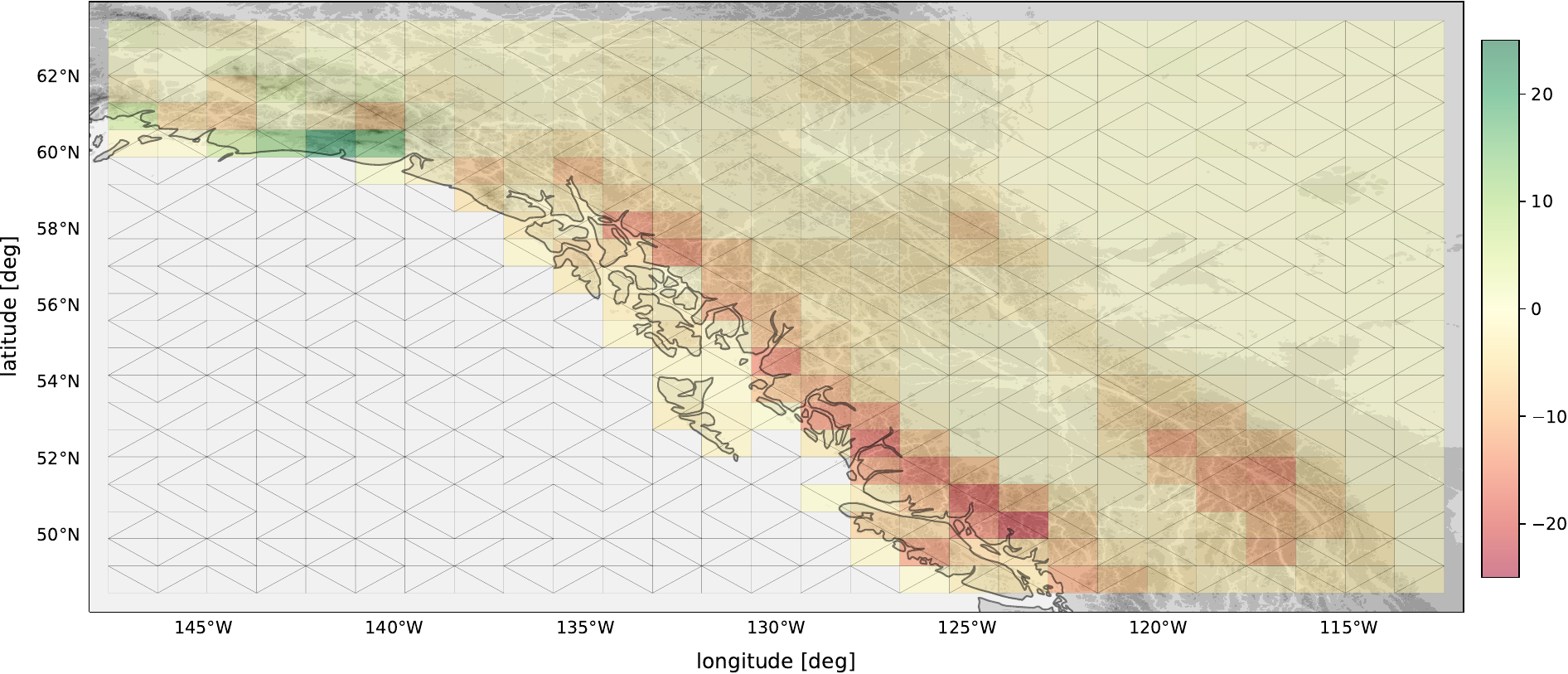}
    \caption{Percentage maximum relative error (MRE) distribution for the finer grid. The MRE distribution is similar to the one in Figure~\ref{fig:error_plot}, although the magnitudes of the MREs are smaller, as can be seen from the color scale. The color bar range is identical to Figure~\ref{fig:error_plot}.}
    \label{fig:error_plot_fine}
\end{figure}

Figure~\ref{fig:error_plot_fine} depicts the MRE distribution over the orography. The distribution of the over- and estimated idealized PMFs remains similar to the error distribution seen in Figure~\ref{fig:percent_mre} even after the grid refinement. This indicates that the method's relative performance is independent of the underlying grid cell size. Instead, the relative error scores are determined by how challenging the underlying orography features are for the CSAM. As noted above, the less intense colors corroborate that the maximum relative errors are less severe than in Figure~\ref{fig:error_plot}, in particular for the underestimated PMFs along the southern mountainous coast.

\subsection{Sensitivity to the background wind}
\label{subsec:sensitivity}
In this section, we investigate two scenarios with the coarser regional study from Section~\ref{subsec:coarse_study}. The first scenario considers a strong background wind speed of $(U,V)=(-40,20)$\,m\,s$^{-1}$. Such a background wind would correspond to a severe storm under real-world conditions. The second scenario studies the effect of the wind direction on the error scores.

\begin{figure}[ht]
    \centering
    \includegraphics[width=1.0\textwidth]{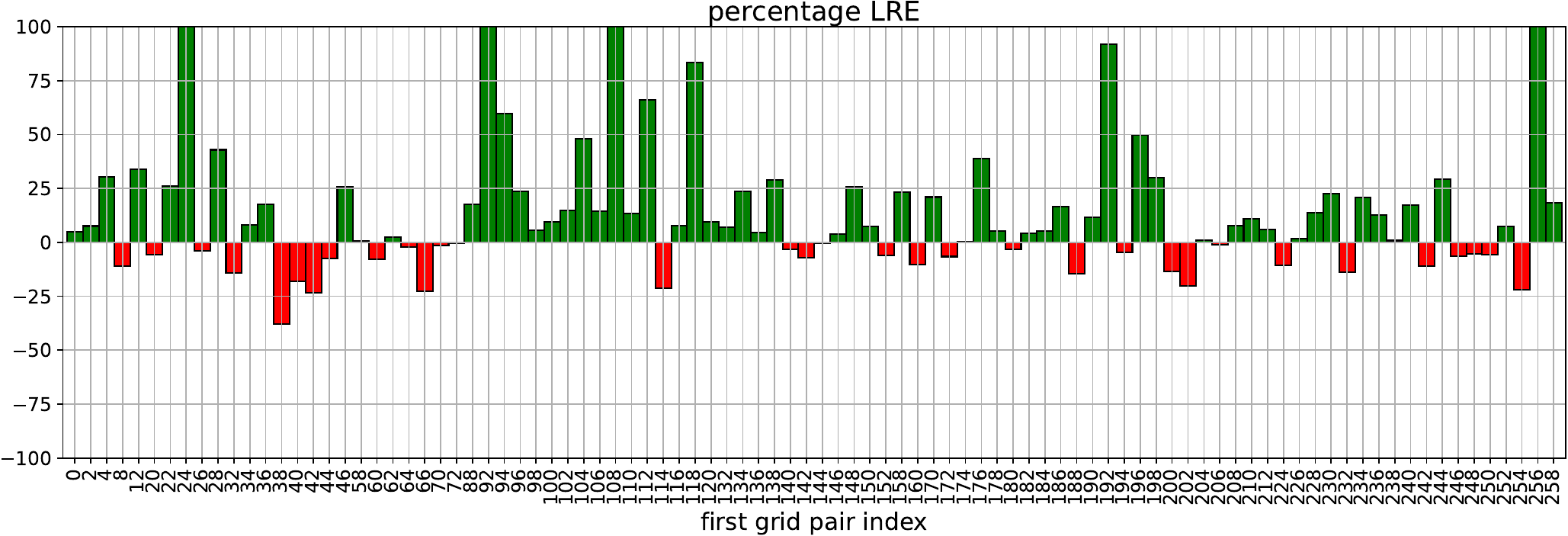}
    \caption{LRE scores for the coarse regional run from Section~\ref{subsec:coarse_study} with strong idealized background wind speed of $(U,V)=(-40,20)$\,m\,s$^{-1}$. Errors beyond $\pm 100\%$ are not depicted. Refer to Figure~\ref{fig:percent_lre} for more details on the LRE scores for the coarse regional run. The average absolute LRE for this strong background wind study is $21.91\%$.}
    \label{fig:high_wind_speed_lre}
\end{figure}

The strong wind speed results in Figure~\ref{fig:high_wind_speed_lre} show a bias towards an overestimation instead of the underestimation bias seen in Figure~\ref{fig:percent_lre}. This result demonstrates that differing wind conditions, e.g., strength and direction, can lead to a substantially different error profile. Nevertheless, the average absolute LRE score in this high wind speed study is 21.91\%, similar to the LRE score of Figure~\ref{fig:percent_lre}. Finally, we mention that the average absolute MRE for this high wind speed case is $4.24\%$. 

\begin{figure}
    \centering
    \includegraphics{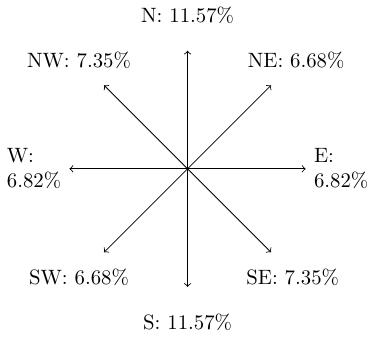}
    \caption{The effect of the wind direction on the MRE scores averaged over all grid cells in the Alaskan Rocky Mountains region. Arrows along the eight cardinal and inter-cardinal directions depict the direction of the wind flow and the associated MRE. The background wind is kept at a constant magnitude of $10$\,m\,s$^{-1}$; see accompanying text for an elaboration.}
    \label{fig:wind_dir}
\end{figure}

Figure~\ref{fig:wind_dir} depicts the average MRE scores over all grid cells for wind blowing from the eight cardinal and intercardinal directions. The magnitude of the background wind is $10$\,m\,s$^{-1}$ in each direction, such that the westerly zonal wind study denoted by `E' is the setup in Section~\ref{subsec:coarse_study}, and $(U,V)=(\sqrt{50}, \sqrt{50})$\,m\,s$^{-1}$ for the intercardinal NE-direction. Two observations can be made from the results in Figure~\ref{fig:wind_dir}. First, we note that the MRE in the `N'- and `S'-directions are the largest, partly due to small local PMF values, which inflate the relative error. The average absolute LREs computed in these two cases are up to 327\%, respectively. Second, the MREs exhibit mirror symmetry along the complimentary cardinal or intercardinal direction, which is to be expected in this idealized setting.

\subsection{Potential biases}
\label{subsec:potential_biases}
Figures~\ref{fig:error_plot} and~\ref{fig:error_plot_fine} reveal a few potential biases of the CSAM method, which will be investigated in this section. We will only consider the coarser R2B4-equivalent grid, as its errors are more pronounced than those on a finer grid. Specifically, the triangle pairs with indices $(200,201)$ and $(24,25)$ will be the foci of our studies, as they present the largest underestimation at $-41.81\%$ and overestimation at $+28.40\%$ in the MRE scores, respectively.

\begin{figure}[ht]
    \centering
    \includegraphics[width=\textwidth]{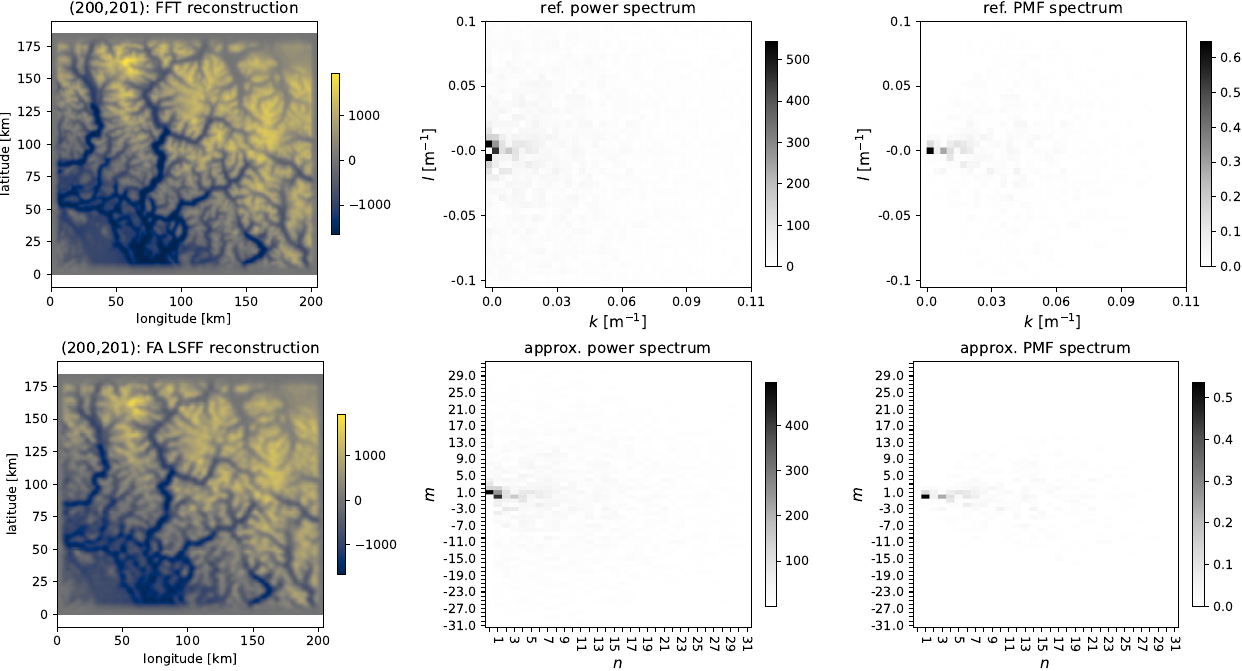}
    \caption{Reconstruction of the physical orography (left column) and power and PMF spectra (middle and right columns) for the quadrilateral grid cell comprising the triangle pair with indices $(200,201)$. (top row) Results obtained from a reference FFT run; (bottom row) results obtained from the FA LSFF run. The top middle and right panels depict the full FFT spectral space equivalent to the physical grid size, while the bottom middle and right panels are the compressed dense FA spectral spaces of $(32 \times 64)$~modes.}
    \label{fig:underestimation_comparison}
\end{figure}

\begin{figure}[ht]
    \centering
    \includegraphics[width=\textwidth]{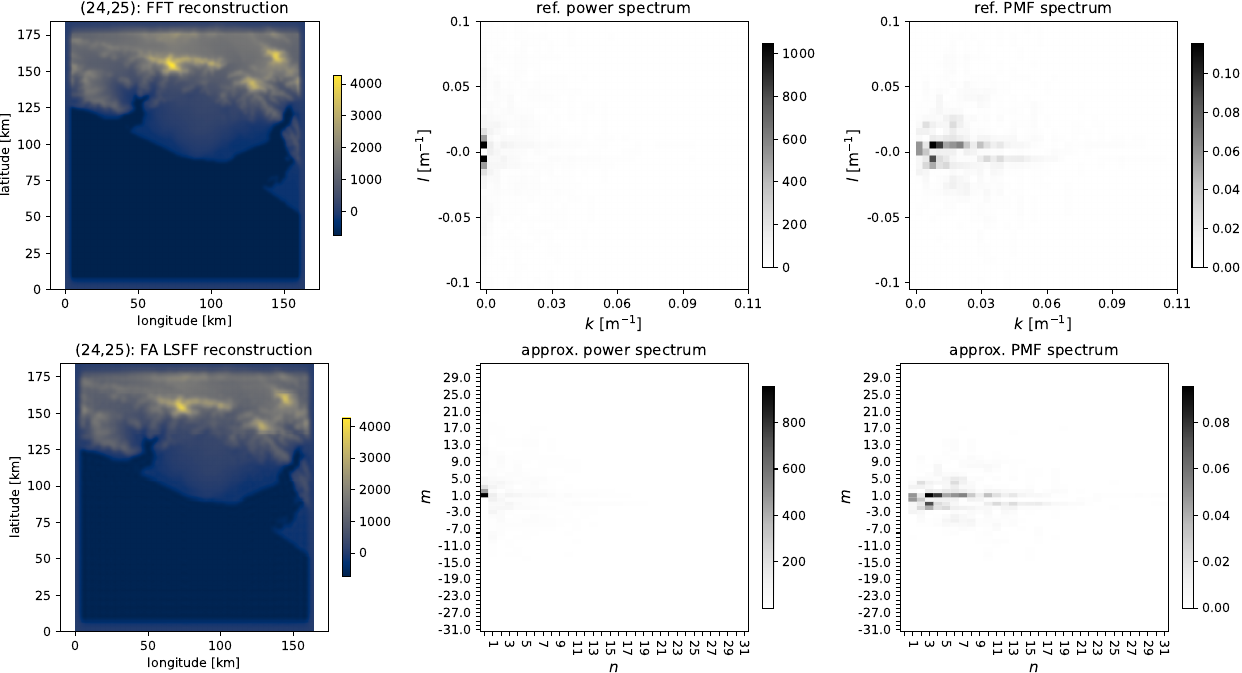}
    \caption{Reconstruction of the physical orography (left column) and power and PMF spectra (middle and right columns) for the quadrilateral grid cell comprising the triangle pair with indices $(24,25)$. (top row) Results obtained from a reference FFT run; (bottom row) results obtained from the FA LSFF run. More details are in the caption of Figure~\ref{fig:underestimation_comparison}.}
    \label{fig:overestimation_comparison}
\end{figure}

Recall that $(\mathcal{N},\mathcal{M})=(32,64)$ is optimal. As such, the FA LSFF on the equidistant quadrilateral grid should give a solution that closely resembles the reference FFT result. This expectation is corroborated by the results shown in Figure~\ref{fig:underestimation_comparison} for the $(200,201)$-pair and Figure~\ref{fig:overestimation_comparison} for the $(24,25)$-pair. Compare the leftmost panels, where the top row depicts the reference FFT solution and the bottom row the FA LSFF approximation. The spectral distributions are largely comparable, recalling that the CSAM does not include the $n=0$, $m<0$ modes; see \eqref{eqn:final_fourier}. The approximated power and PMF spectra are largely similar (compare the top panels to the bottom panels for the middle and right columns in Figures~\ref{fig:underestimation_comparison} and~\ref{fig:overestimation_comparison}). Note that in all the figures of this section, the padded region where the tapering occurs is depicted in the physical reconstruction plots (left columns).

\begin{figure}[ht]
    \centering
    \includegraphics[width=\textwidth]{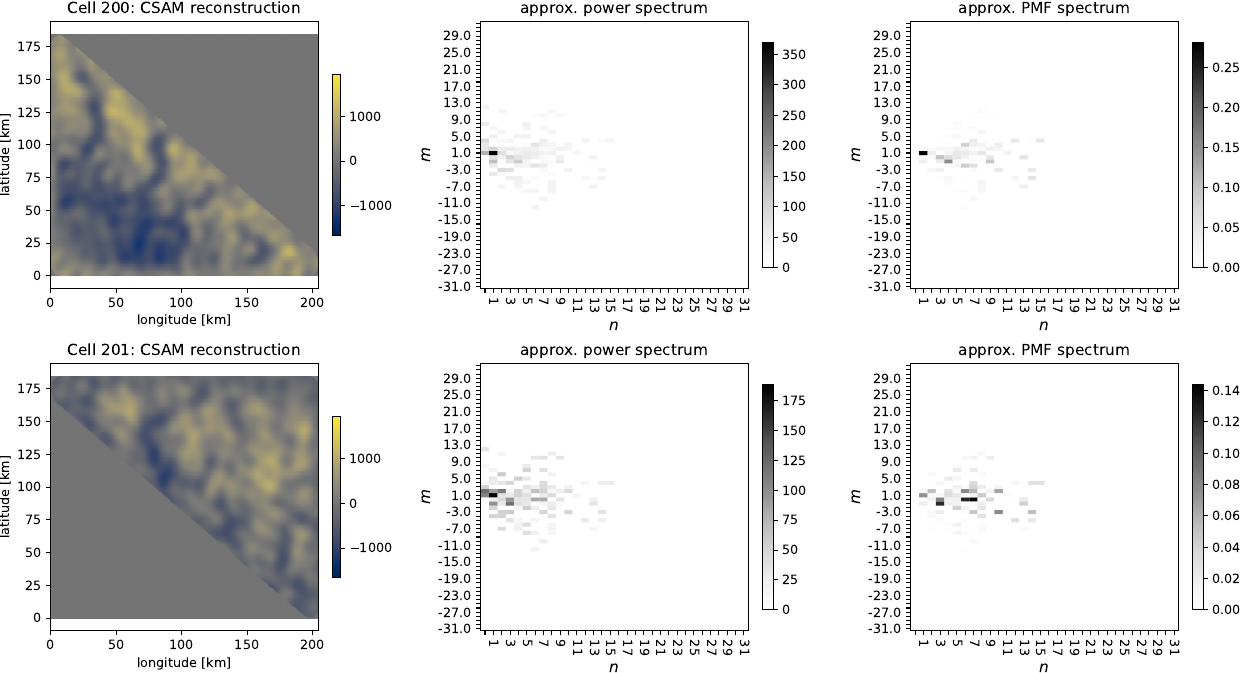}
    \caption{Reconstruction of the physical orography (left column) and power and PMF spectra (middle and right columns) for the non-quadrilateral grid cell with the index $200$ (top row) and $201$ (bottom row). The middle and right panels depict the sparse compressed spectral space with at most $N_\text{modes}$ number of non-zero spectral modes.}
    \label{fig:underestimation_csam}
\end{figure}

\begin{figure}[ht]
    \centering
    \includegraphics[width=\textwidth]{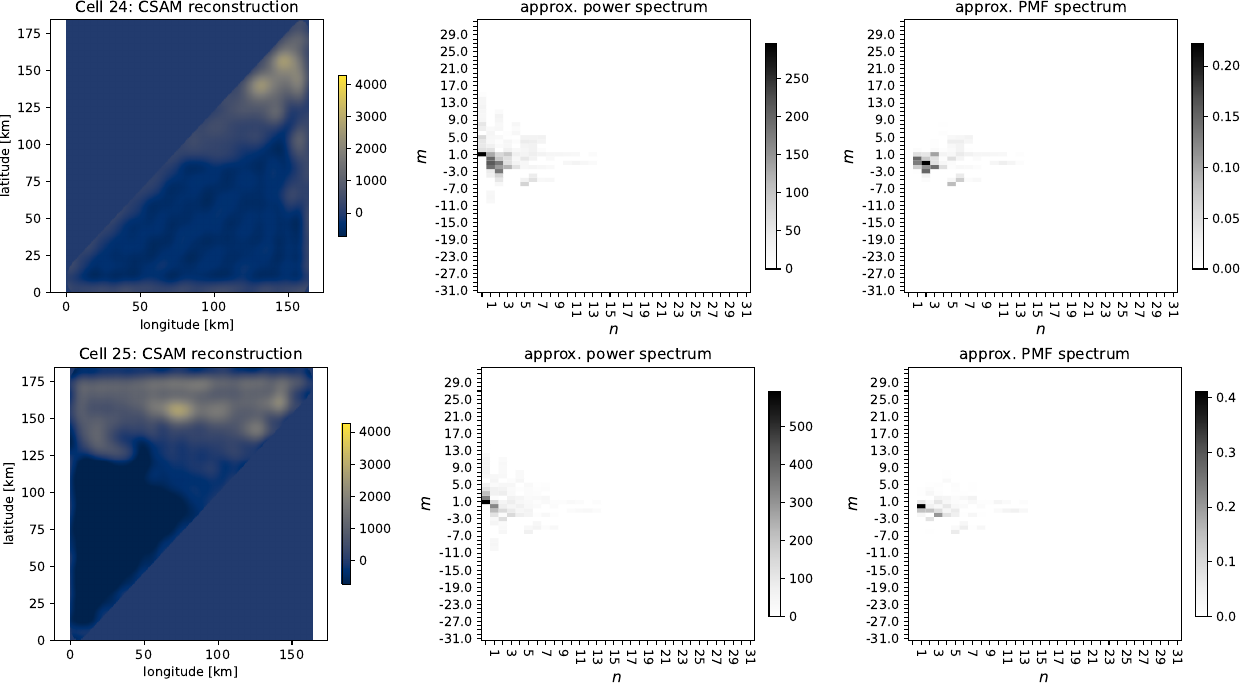}
    \caption{Reconstruction of the physical orography (left column) and power and PMF spectra (middle and right columns) for the non-quadrilateral grid cell with the index $24$ (top row) and $25$ (bottom row). The middle and right panels depict the sparse compressed spectral space with at most $N_\text{modes}$ number of non-zero spectral modes.}
    \label{fig:overrestimation_csam}
\end{figure}

As long as quadrilateral grid cells are concerned, we are not faced with any unpleasant surprises. However, the performance of the CSAM deteriorates in the approximation of non-equidistant data points in a non-quadrilateral grid cell. The reconstructed physical orography and the spectra of the CSAM approximations are depicted in Figure~\ref{fig:underestimation_csam} for the $(200,201)$-pair and Figure~\ref{fig:overrestimation_csam} for the $(24,25)$-pair. In both these cases, the physical orography recovered is no longer as close to the reference FFT-recovered orography, and small-scale information is lost. In addition to the non-quadrilateral grid cell, this loss of information is due to the further compression of the spectral space down from $(32 \times 64)$ to $100$ sparse modes and the non-equidistant topographic data points used in the SA LSFF. Nevertheless, large-scale features are qualitatively preserved. Elaboration on the potential biases in both cases will be provided separately below.

The orography in $(200,201)$ is complex with many fine-scale dendritic features. The CSAM is based on linear regression, such that large-scale features are fitted best, and the fitting for increasingly smaller features progressively deteriorates. From its setup, the method would struggle to spectrally encode the information of complex orographies. Indeed, the FA LSFF result in Figure~\ref{fig:underestimation_comparison} shows superb physical similarities to the reference orography, but even then, the idealized PMF exhibits a 20.30\% underestimation from $\mathcal{P}^{\text{ref}}$. Comparing the approximated CSAM spectra in Figure~\ref{fig:underestimation_csam} to the FA LSFF spectrum in Figure~\ref{fig:underestimation_comparison}, we notice that the maximum amplitudes are smaller in the CSAM case, and this ``loss'' in power is not made up by more numerous large-amplitude spectral modes. This explains the considerably larger underestimation of the CSAM-approximated PMF compared to the FA LSFF case (41.82\% against 20.30\%). The cluster of grid cells around $(200,201)$ with significant underestimation errors feature similar dendritic orographic structures over large changes in altitude.

Observe that the power and PMF spectra approximated from the non-quadrilateral orography features in Figure~\ref{fig:underestimation_csam} no longer closely resemble the reference spectra. This is expected as the underlying orographic features are now vastly different. For example, the dominant spectral mode identified in the quadrilateral FA LSFF is $(n,m)=(1,0)$ (middle panels of Figure~\ref{fig:underestimation_comparison}), which could be qualitatively verified by simply observing the physical orography. However, in both the non-quadrilateral approximations, the dominant modes identified are $(n,m)=(1,1)$ (middle panels of Figure~\ref{fig:underestimation_csam}), which again may be verified why this is the case from a quick visual inspection of the non-quadrilateral orographies. Therefore, we emphasize that the CSAM is not meant to reproduce the spectral distribution of the orographic features in a quadrilateral grid cell accurately, as this would be impossible. Instead, the goal is to produce from the respective approximated spectra an estimate of the wave-action density (and the PMF) that is as close to the physical value as possible.

The overestimation for the $(24,25)$-pair is due to the drastic changes to the underlying orography in the non-quadrilateral grids. For the quadrilateral grid cell depicted in Figure~\ref{fig:overestimation_comparison}, the dominant spectral mode is $(n,m)=(0,\pm 1)$ as verifiable from a quick visual inspection of the physical orographies (left panels) and the power spectra (middle panels) in the figure. However, for an idealized background wind of $(U,V)=(10,0)$\,m\,s$^{-1}$, these modes do not contribute to the computed PMF. Instead, the PMF contributions come predominantly from spectral modes with smaller amplitudes; see the rightmost panels. In the non-quadrilateral decomposition shown in Figure~\ref{fig:overrestimation_csam}, the dominant spectral mode is no longer only $(n,m)=(0,1)$, especially for cell index 24 (top row, middle panel). In these cases, spurious spectral modes with $n>0$ have large, non-zero amplitudes, and these modes contribute substantially to the idealized PMF value. Note that the maximum spectral amplitudes are nevertheless smaller than in the quadrilateral case, as they should be; compare, e.g., the color bars of the middle panels Figure~\ref{fig:overrestimation_csam} to~\ref{fig:overestimation_comparison}. However, the difference to the underestimation case discussed above is that more spectral modes with significant contribution to the PMF for the given background wind are triggered in the CSAM-approximated spectra. These spurious modes may be characterized as unwanted spectral noise.

The two potential sources of biases that led to the most significant errors in the regional study can be summarized as follows. Complex orographic features, such as the sharp valleys and gorges seen in the coastal mountainous terrain in and around the grid pair $(200,201)$, are a challenge for the linear regression-based CSAM and lead to underestimation of the SSO spectrum. On the other hand, if the non-quadrilateral decomposition substantially changes the underlying orographic features so that the dominant modes are no longer strongly identifiable, then spurious spectral modes with small amplitudes will become dominant in the approximated spectrum. This leads to a potential overestimation of the wave-action density under certain wind conditions.

\subsection{Iterative refinement: Selected results}
\label{subsec:iterative_refinement_results}
The potential biases of the CSAM are unavoidable, given the external constraints. However, suppose the reference target spectrum is available, such as the idealized Delaunay triangulation tests we have set up above. In that case, we may improve, in a deterministic fashion, the error scores via the iterative refinement extension from Section~\ref{subsec:it_refinement}. By deterministic, we mean that the method's principal capabilities and behavior can be characterized in a predictable manner. We will apply the iterative refinement extension to a collection of the nine worst MRE offenders from the coarse regional study in Section~\ref{subsec:coarse_study}, namely the following grid cell pairs,
\begin{equation}
     (24,25); (42,43); (92,93); (152,153); (160,161); (180,181); (200,201); (202,203); (238,239).
     \label{eqn:cell_indices}
\end{equation}
The error tolerance used is LRE$=\pm20\%$, which means we repeat the IR process until the LRE is within this bound. 

An additional modification is made to the experimental parameters from Section~\ref{subsec:coarse_study}, such that $(\mathcal{N},\mathcal{M})=(16,32)$ is used in this IR study. The reasoning for this is as follows. The IR extension works by scale separation, either by progressively including more small-scale features that the CSAM did not adequately capture or by progressively filtering spurious small-scale noise introduced in the approximated spectrum. However, there is no meaningful way to separate the residual features of the reconstructed orography on the non-quadrilateral cells from the reference on a quadrilateral cell. For example, one major hurdle in this regard would be the discontinuity along the diagonal face of the Delaunay triangles if one were to sum the two reconstructed orographies. Instead, the IR algorithm uses the next-best quadrilateral approximation available to compute the residual orographic features, i.e., the reconstructed orography from the FA step. However, as we have seen in Figures~\ref{fig:underestimation_comparison} and~\ref{fig:overestimation_comparison}, the physically reconstructed orography on the quadrilateral grid cell can strikingly resemble the reference orography, particularly when the choice of $(\mathcal{N},\mathcal{M})$ is optimal. In such cases, taking the residual orography in \eqref{eqn:phys_residual} no longer makes meaningful sense as a correction target, as the residual no longer resembles the error present in the non-quadrilateral reconstructions. Therefore, we circumvent this issue by artificially introducing an additional error between the CSAM and the target through a suboptimal choice of the dense spectral domain, ensuring we always get a meaningful correction target in the IR procedure. The other CSAM free parameters are not modified in the IR process.

\begin{figure}[ht]
    \centering
    \includegraphics[width=\textwidth]{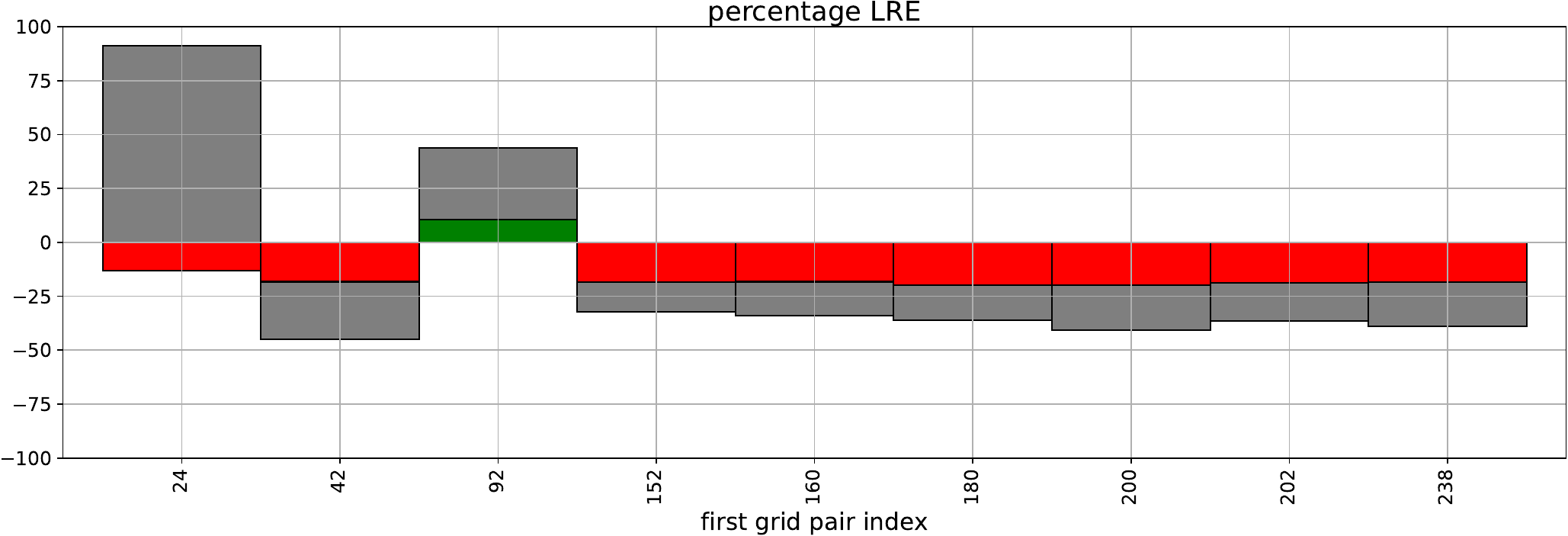}
    \caption{LRE scores for the nine grid-cell pairs with the worst MRE scores from the coarse regional study before and after the iterative refinement process; see \eqref{eqn:cell_indices} for the cell indices. Gray bars denote the initial LREs, and the green and red colored bars are the final LREs after the IR procedure with convergence tolerance of $\pm20\%$. Note that the FA spectral space size is $(\mathcal{N},\mathcal{M})=(16,32)$; refer to the accompanying explanation for more details. See Figure~\ref{fig:error_plot} for more details on the LRE scores.}
    \label{fig:ir_results}
\end{figure}

The final IR LRE scores for the nine grid cells are shown in Figure~\ref{fig:ir_results}. The gray colored bars are the initial LREs, and note that they differ from the ones in Figure~\ref{fig:percent_lre} due to the smaller dense spectral space used in the FA. The red and green bars are the final LREs obtained after the convergence of the IR procedure, and the mean absolute LRE is reduced from $44.28\%$ to $17.32\%$ via iterative refinement. However, this encouraging result is not illustrative of the IR process. Below, we establish precisely what the IR accomplishes.

\begin{figure}[ht]
    \centering
    \includegraphics[width=\textwidth]{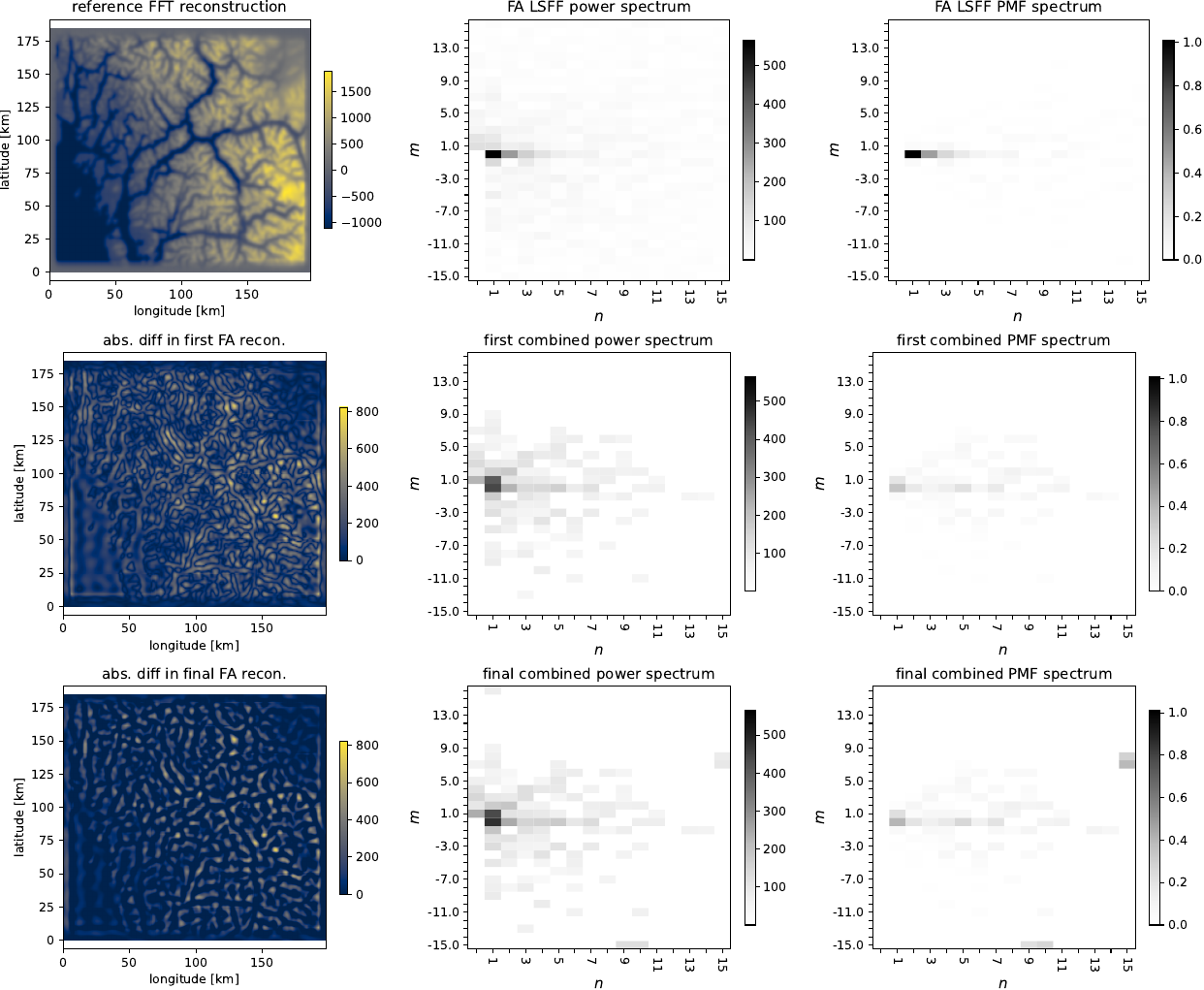}
    \caption{Iterative refinement process applied to grid pair (42,43). (top left panel) Reference reconstructed orography from the FFT process. (left column, center and bottom panels) Absolute difference of the reconstructed orography from the reference before and after the IR process, respectively. (middle column, from top to bottom) Power spectrum after the FA step; combined CSAM-approximated spectrum from the two non-quadrilateral grid cell before the IR process; and after the IR process. (right column, from top to bottom) PMF spectra with identical configurations to the middle column. Except for the top left panel, all color bars are column-wise identical.}
    \label{fig:iter_plots}
\end{figure}

Let us focus on the grid pair (42,43), which exhibits the most significant underestimation error. The top left panel of Figure~\ref{fig:iter_plots} depicts the reference FFT reconstruction of the original orography, and the top middle and top right panels are the power and PMF spectra from the FA step. The latter two panels are close to their respective FFT spectrum and may be taken as the reference solutions (comparison not shown). The leftmost panel of the center row depicts the absolute difference between the reference and FA reconstructions after the first CSAM FA step and before invoking the IR process. The panel below (bottom row, leftmost panel) depicts this difference after the last iteration of the IR process. Fourteen iterations were required to reduce the error from an initial -45.10\% to -18.26\% after convergence. These two leftmost panels represent the error in the reconstructed physical orographies before and after applying IR. The sum of the CSAM-approximated spectra from the two non-quadrilateral grid cells is termed the ``combined spectrum''. The middle and right panels of the center row are combined power and PMF spectra after the first CSAM FA step and before the IR process. The two panels below them are obtained after the final IR iteration.

Before the IR process (center row of Figure~\ref{fig:iter_plots}), the error in the physical orography is significant (left column), and the power spectrum (middle panel) exhibits a similar distribution in the spectral modes as the reference spectrum. However, the degrees of freedom available to the CSAM-approximated spectrum are about five times smaller than the dense FA-approximated spectrum, and each sparsely distributed CSAM spectral mode carries a larger amplitude than its FA LSFF counterpart, compare the middle panels in the top and middle rows. As this is the combined spectrum, the amplitude contributions are from two separate non-quadrilateral grid cells. Therefore, the combined PMF spectrum (right panel) has a significantly smaller amplitude for its most dominant modes around $(n,m)=(1,-1)$. This underestimation of the dominant amplitudes is somewhat made up by a more extensive distribution of modes with significant amplitudes (gray patches in the rightmost panel).

Compare the results before the IR process to the ones obtained after the IR process (middle and bottom rows of Figure~\ref{fig:iter_plots}). The power spectrum (middle panel) looks similar to the one before the IR process except for a few significant modes at the boundary of the spectral domain, e.g., $(n,m)=(9,-15)$ or $(n,m)=(15,7)$. Regarding the PMF spectrum (right panel), the dominant $(1,-1)$-mode has a slightly larger amplitude, and the newly introduced modes at the boundary of the spectral domain have a substantial contribution to the PMF. At first glance, these high-frequency modes seem to be spurious. However, recall that the CSAM compresses the spectral representation down to a hundred sparsely distributed modes, and each iteration step selects the hundred most prominent modes from the spectra before and after refinement. This means that the large-amplitude small-scale modes that survive the iteration process are the ones that encode the small-scale physical features as best as possible with as few modes as possible. I.e., these high-frequency modes are the best compromises to represent small-scale orographic features given the constrained spectral space.

The observation made in the previous paragraph is corroborated by the improvement in the absolute difference of the final reconstructed orography from the reference orography (bottom row, leftmost panel of Figure~\ref{fig:iter_plots}). Quantitatively, the IR process reduces the relative $L_2$-error in the reconstructed physical FA orographies from 1.02 to 0.97. Thus, including the large-amplitude small-scale modes in the approximated spectrum improves the representation of the SSO and yields better PMF error scores. Given the limited number of spectral modes afforded to the method, these high-frequency modes may be considered a judicious spectral clustering of important small-scale information by the CSAM; this observation will be further discussed in Section~\ref{sec:discussions}.

These results verify two crucial aspects of the CSAM. First, the CSAM is deterministic, and given a reference target, the error scores can be minimized via an iterative procedure. Second, the sources of errors in the method can be methodically characterized and removed by a progressive spectral separation of scales. So even though such an IR process is only applicable in very specific cases where the reference spectrum is known, implementing and investing it is crucial in establishing the principal capabilities and behavior of the CSAM.

Two further remarks on the IR extension are necessary. First, the convergence of the IR extension depends on the choice of the correction target, and the smaller the error that the initial correction target provides, the greater the number of iterations the IR will need to achieve the desired error tolerance. Therefore, choosing a smaller spectral space such as the one used in this section may be meaningful, e.g., with half the optimal number of spectral modes in both directions. Second, the LRE is a better metric in the IR study, as we are interested in answering how well the error in each grid cell can be improved via such an iterative procedure. Therefore, we omit showing the MRE scores. 


\section{Discussions}
\label{sec:discussions}
The spectrum of orographic features represented in a non-quadrilateral grid cell inherently differs from that in the encompassing quadrilateral domain. Nevertheless, the CSAM can decently approximate the pseudo-momentum fluxes over a region featuring different terrain types. It is ultimately this ability to faithfully reproduce physically relevant quantities for which the CSAM is developed, and the principal capabilities and behavior of the method have been established through numerical studies. The $\pm10$\% maximum relative error we encountered in the regional study over a coarse grid suitable for climate simulations substantially improves state-of-the-art operational SSO representations. This is notwithstanding the method's robustness to handle non-equidistant data points in a non-quadrilateral grid cell and more than two orders-of-magnitude compression of the problem's dimensions while featuring scale awareness.

Moreover, the numerical experiments involving real-world terrain data were all conducted with the same free parameters choice across different grid cell sizes and background wind speeds. When changes to the free parameters were made, e.g., with the choice of the optimal $(\mathcal{N},\mathcal{M})$ values or the halving of the dense spectral domain in the iterative refinement experiments of Section~\ref{subsec:iterative_refinement_results}, these changes were motivated by either physical or algorithmic arguments. A potential user of the CSAM may be guided to make a judicious first guess about what these free parameters could be, and this first guess is likely to work without requiring further tuning. Hence, it is worth emphasizing again that the free parameters are not to be seen as tuning parameters, and the CSAM does not require the computationally expensive and time-consuming process of tuning the method to work with different configurations of an underlying grid or orographic dataset.

In the least-squares Fourier fitting of Section~\ref{subsec:lsff}, the deliberate choice to use an $L_2$-regularizer was made. An $L_1$-regularizer would shrink the spectral SSO representation as intended from the second approximation step, but this would not explicitly allow the user to select the number of spectral modes. Moreover, the $L_1$-regularized spectrum does not generally lead to better error scores for a fixed $\lambda_{\text{SA}}$ value. Therefore, we retain the choice of an $L_2$-regularizer so that the choice of $N_{\text{modes}}$ is a free parameter, but this regularizer may be modified depending on the application of the CSAM.

Another consideration is using either a fast Fourier transform or least-squares Fourier fitting for the first approximation step in the CSAM algorithm. This choice was briefly mentioned in Section~\ref{subsec:intro_constraints}. While least-squares Fourier fitting yields better error scores (see \ref{apx:fft_vs_lsff} for more details), this choice may not always be appropriate. An issue with the parameterization of orographic gravity waves is the blocked layer depth, below which gravity waves are not launched but break down into small-scale turbulence \cite{lott1997new}. Characterizing this blocked layer depth and rescaling the effective orography's amplitudes are fields of active research \cite{van2021towards}. Matters are further complicated because the blocked layer depth depends on the background wind speed and orientation. As such, online recomputation of the effective mountain height in an SSO representation depending on the background winds and orientation may be a worthy pursuit, albeit computationally expensive, to investigate the susceptibility of orography gravity wave parametrization to these factors. In answering this question, the fast Fourier transform would be a sensible compromise to achieve considerable computational speed-up.

The deliberate choice of not interpolating the gridded non-equidistant data on the right-hand side of the least-squares Fourier fitting in the second approximation step of the algorithm introduces additional errors. However, reasonable error scores can be achieved by taking the mean distance between the topographic data points, as seen in the results presented in Section~\ref{sec:results}. A major issue may arise at the geographical poles when the topographic data becomes strongly non-equidistant, and the validity of the mean-distance assumption no longer holds generally. Therefore, interpolating the non-quadrilateral topographic data in the second approximation may be used to achieve better error scores and may be necessary under these circumstances. However, we consider such a study beyond the scope of this work, as one of our goals is to establish the CSAM's robustness under the more challenging scenario featuring non-equidistant data points.

The CSAM is a method to approximate the spectrum of features given some constraints, and as such, it naturally suffers from potential biases that limit the method's effectiveness. These approximation errors propagate through the computation of the physically relevant quantities and as feedback into the dynamical core. However, this error propagation is not unique to the CSAM, and the error scores may be improved with more careful treatment of the SSO in a preprocessing step. Nevertheless, the error scores are already relatively reasonable, as shown in the numerical studies of Section~\ref{sec:results}. The strength of the CSAM is its robustness and deterministic behavior, which is achievable with a relatively straightforward implementation and setup.

The number of ray-volumes launched in gravity-wave raytracers is limited by computational requirements. To this end, an alternative to the coarse-graining of spectral space mentioned in Section~\ref{subsec:potential_variations} is to take all spectral components but randomly pick a subset of these components to launch at each time step \cite{de2015parameterization, ribstein2022can}. This approach is similar to Monte Carlo-type simulations. The CSAM naturally allows for such a stochastic parametrization of orographic gravity waves. However, a more physical approach may be the selective clustering of relevant spectral components. We have seen its effectiveness as a compromise to represent small-scale orographic features in the results presented in Figure~\ref{fig:iter_plots} and the accompanying explanation. Suppose the CSAM identifies a cluster of relevant modes despite the sparse spectral approximation. In that case, these modes can be clustered and represented by one mode that represents the average of its neighbors, further reducing the complexity of the problem. This investigation will be left to a future study after the CSAM is coupled to the MS-GWaM raytracer and the ICON dynamical core. 

The iterative refinement extension to the CSAM introduced in Sections~\ref{subsec:it_refinement} and~\ref{subsec:iterative_refinement_results} demonstrates that the approximation of the spectrum could be improved given a sensible optimization target. The idealized experiments studied above are set up so this target is readily available. However, for real-world applications, the non-quadrilateral decomposition of the global grid may not have a simple quadrilateral counterpart. In such scenarios, the optimization target could be a relevant physical quantity, such as the pseudo-momentum fluxes, obtained from reanalysis or observational data. Given these optimization targets, the a posteriori iterative refinement would be possible to produce the best approximation of the subgrid-scale orographic spectrum. However, a challenge is the separation of pseudo-momentum fluxes arising from orographic sources only, and this is an ongoing research question; see, e.g., \citeA{prochazkova2023sensitivity}.

This work primarily focuses on developing the CSAM to produce a subgrid-scale orographic spectrum as an input to a gravity-wave raytracer. However, the CSAM developed here is general and can be used for various more generic applications. For example, the iterative refinement procedure for spectral analysis yields scale separation and selection for two-dimensional non-equidistant data points in a non-quadrilateral domain. The method could also approximate periodograms in time series analysis that produces physically sound spectral amplitudes (recall from Section~\ref{subsec:idealized_tests} that the CSAM beats the error scores of the least-squares spectral analysis even in idealized tests). Finally, the CSAM may find a broader range of applications if extended to work with less structured, non-gridded data. This extension will be left to future work.

\section{Conclusions}
\label{sec:conclusions}
This work introduces a novel spectral analysis method for gridded, non-equidistant data points distributed in a non-quadrilateral grid cell. The energy of the spectrum approximated by the method is physically sound. It can be used as a scale-aware representation of subgrid-scale orography in real-world numerical weather prediction independent of the grid cell size of the dynamical core. The resultant error is within $\pm 10\%$ relative to the most considerable pseudo-momentum flux for the regional study with real-world topographic data investigated here.  

Furthermore, as a physical parametrization of orographic gravity waves can be computationally expensive, the constrained spectral approximation method introduced here can achieve more than two orders of magnitude reduction of the problem's complexity by representing the subgrid-scale orography with fewer than a hundred spectral modes. The method is robust to the choice of free parameters used, and tuning is generally unnecessary even when external conditions change, such as the grid cell size and background wind. Implementing this least-squares spectral analysis-based method is straightforward and does not require extensive effort to set up and use.

The method will be required for initializing the input for an orographic gravity wave source to a gravity-wave raytracer or other spectral-based orographic gravity-wave parametrization schemes coupled to a dynamical core with a geodesic grid. The constrained spectral approximation method might also provide an improved alternative to existing approximations of subgrid-scale orography. Beyond these use cases, the method could be used for spectral analysis, scale separation and selection, and data compression. The method may find potential applications outside of weather forecasting.

%


%
\appendix
\section{Availability Statement}
The simulation datasets generated in this work have been deposited in a Zenodo repository and may be accessed via \citeA{datasets}. Input parameters necessary for reproducing the simulation runs are included as metadata to the datasets.

Furthermore, the version of the CSAM source code utilized for conducting the simulations has been released on GitHub \cite{pyCSAM}. The source code includes detailed documentation and instructions for running the simulations.

\section{Derivation of Eq.~\ref{eqn:final_fourier}}
\label{apx:final_fourier_deriv}
We start with a truncated Fourier representation of the physical orography,
\begin{equation}
    h(x,y) = \sum_{n=-\mathcal{N}}^\mathcal{N} \sum_{m=-\mathcal{M}}^\mathcal{M} \hat{h}_{n,m} \, e^{i( k_n x + l_m y )}.
    \label{eqn:fourier_rep}
\end{equation}
following the notations used in Section~\ref{subsec:lsff}. Expanding \eqref{eqn:fourier_rep} in terms of the real and complex quadrants, we obtain
\begin{align}
    h(x,y) &= \hat{h}_{0,0} + \sum_{\substack{n=-\mathcal{N}\\n \neq 0}}^\mathcal{N} \hat{h}_{n,0} \, e^{ik_n x} + \sum_{\substack{m=-\mathcal{M}\\m \neq 0}}^\mathcal{M} \hat{h}_{0,m} \, e^{il_m y} 
    \nonumber \\ 
    &\qquad  + \left( \sum_{n=-\mathcal{N}}^{-1} \sum_{m=-\mathcal{M}}^{-1} + \sum_{n=-\mathcal{N}}^{-1} \sum_{m=1}^{\mathcal{M}} + \sum_{n=1}^{\mathcal{N}} \sum_{m=-\mathcal{M}}^{-1} + \sum_{n=1}^{\mathcal{N}} \sum_{m=1}^{\mathcal{M}} \right) \hat{h}_{n,m} \, e^{i(k_n x + l_m y)}
    \nonumber \\
    &= \hat{h}_{0,0} + \sum_{n=1}^\mathcal{N} \left( \hat{h}_{n,0} \, e^{ik_n x} + \hat{h}_{n,0}^\ast \, e^{-ik_n x}\right)  + \sum_{m=1}^\mathcal{M} \left( \hat{h}_{0,m} \, e^{il_m y} + \hat{h}_{0,m}^\ast \, e^{-il_m y} \right)
    \nonumber \\ 
    &\qquad  +  \left( \sum_{n=1}^{\mathcal{N}} \sum_{m=1}^{\mathcal{M}} + \sum_{n=1}^{\mathcal{N}} \sum_{m=-\mathcal{M}}^{-1} \right) \left[ \hat{h}_{n,m} \, e^{i(k_n x + l_m y)} + \hat{h}_{n,m}^\ast \, e^{-i(k_n x + l_m y)} \right],
    \label{eqn:fourier_rep_intermediate}
\end{align}
where since $h(x,y)$ is real, the negative of a Fourier component is its complex conjugate, e.g., $\hat{h}_{-n,0} = \hat{h}^\ast_{n,0}$. Now, expanding \eqref{eqn:fourier_rep_intermediate} in terms of its real and imaginary components, i.e., $\hat{h}_{n,m} = \hat{h}^{(r)}_{n,m} + \hat{h}^{(i)}_{n,m}$ and $e^{i(k_n x + l_m y)} = \cos(k_n x + l_m y) + i \sin (k_n x + l_m y)$, yields
\begin{align}
    h(x,y) = \hat{h}_{0,0} &+ \sum_{n=1}^\mathcal{N} 2 \left[ \hat{h}_{n,0}^{(r)} \cos(k_n x) - \hat{h}_{n,0}^{(i)} \sin(k_n x) \right]
    \nonumber \\ 
    & + \sum_{m=1}^\mathcal{M} 2 \left[ \hat{h}_{0,m}^{(r)} \cos(l_m y) - \hat{h}_{0,m}^{(i)} \sin(l_m y) \right]
    \nonumber \\ 
    & +  \sum_{n=1}^{\mathcal{N}} \sum_{\substack{m=-\mathcal{M}\\m \neq 0}}^\mathcal{M} 2 \left[ \hat{h}_{n,m}^{(r)} \cos(k_n x + l_m y) - \hat{h}_{n,m}^{(i)} \sin(k_n x + l_m y) \right],
\end{align}
which may be written compactly in the form of \eqref{eqn:final_fourier}.

\section{FFT vs LSFF in the First Approximation Step}
\label{apx:fft_vs_lsff}
\begin{figure}[ht]
    \centering
    \includegraphics[width=1.0\textwidth]{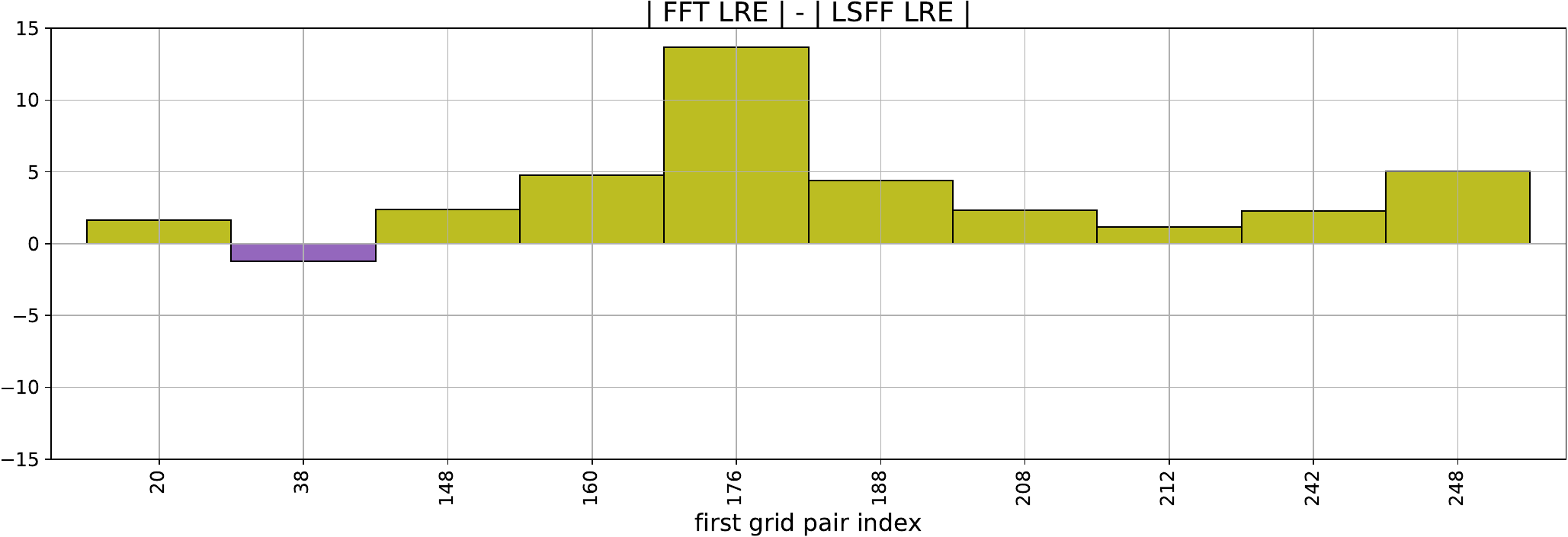}
    \caption{Comparison of the choice of FFT vs LSFF for the FA step of the CSAM algorithm, i.e., the difference of the absolute LSFF LRE from the absolute FFT LRE. Positive differences (yellow bars) correspond to the LSFF outperforming FFT and negative differences otherwise (purple bars). The grid cell pairs are randomly chosen; see Figure~\ref{fig:delaunay} for more details. Note that the vertical axis extent is from $[-15,15]\%$ LRE difference.}
    \label{fig:dfft_vs_lsff}
\end{figure}

This appendix briefly states a few results comparing the choice of fast Fourier transform against the least-squares Fourier fitting of Section~\ref{subsec:lsff} in the first approximation step of the CSAM algorithm. This study is conducted over ten randomly selected grid cell pairs, with the indices highlighted in Figure~\ref{fig:delaunay}. All simulation parameters are identical to the ones used in Section~\ref{subsec:coarse_study}. The difference of the absolute LSFF LRE from the absolute FFT LRE for each of these grid cells is depicted in Figure~\ref{fig:dfft_vs_lsff}, and positive values (yellow bars) mean that LSFF outperforms FFT and vice versa (purple bars).

When FFT outperforms LSFF (purple bars), the improvement is marginal, within two percentage points. However, the LSFF provides substantial advantages for grid cells with fine-scale orographic features. Recall that by choosing FFT in FA, the CSAM only compresses the dimension of the spectral modes once. Small-scale features identified in the FFT spectrum via many small amplitude modes cannot be accurately reproduced when the sparse spectral distribution is selected in the SA step. On the other hand, the choice of LSFF in FA already compresses the spectrum to a dense subspace, and this step requires the algorithm to pre-select the best approximation of these small features. Similar to the observations made in Section~\ref{subsec:iterative_refinement_results}, these high-frequency modes may carry substantial amplitude and become relevant in the SA step. Overall, the average absolute LRE for the LSFF run is 25.5\% versus 29.20\% for the FFT run. The MREs do not have a meaningful notion for a randomly selected non-neighboring set of grid cells, and these results are omitted here. 

The approximately 4.0\% improvement in LRE scores may be negligible when computational speed becomes a requirement, e.g., in online investigations of the blocked layer depth (see the discussions in Section~\ref{sec:discussions}). Indeed, the FFT FA choice is more than twice as fast as the LSFF FA in terms of computational efficiency. The compute time of the CSAM over the ten grid cells is 27.41\,s against 69.15\,s, respectively. Due to the vast difference in this performance, the FFT FA choice is preferred whenever computational efficiency matters.





\section{Tapering of the topographic data}
\label{apx:tapering}
A Shapiro-like filter tapers the boundary of the topography data \cite{shapiro1970smoothing}, where the significant modification here is the dampening of high-frequency features only occurs along the edges of the possibly non-quadrilateral domain of interest. This is achieved by artificially diffusing the Boolean mask that selects for points within the non-quadrilateral domain. Note that the diffused mask will naturally no longer contain only Boolean values. Specifically, we solve the diffusion equation
\begin{equation}
    \label{eqn:artificial_diffusion}
    \partial_\tau u(\tau, x, y) = \Delta u(\tau, x, y),
\end{equation}
where $\tau$ is an artificial time we introduce, and $u$ is the mask such that $u(0,x,y)=1$ for data points inside the domain to be tapered and $0$ otherwise. Discretization of \eqref{eqn:artificial_diffusion} leads to
\begin{equation}
    u^{r+1} = u^r + \Delta \tau \, \widetilde{\Delta} u^r,
    \label{eqn:discrete_artificial_diffusion}
\end{equation}
where we suppressed the arguments of $u$, and $r$ indexes the discrete artificial time steps. $\Delta \tau$ is the artificial time-step size and $\widetilde{\Delta}$ is a discretization of the Laplace operator via the nine-point \citeA{oono1987computationally} stencil. After each time integration step \eqref{eqn:discrete_artificial_diffusion}, the values inside the domain to be tapered are reset to 1's. Physically, this is akin to diffusing an infinite source, and this is done to ensure that no information inside the domain of interest is lost in the diffusion process.

Implementing this tapering choice is straightforward yet applicable to any domain structure. The free parameter choices are $\Delta \tau$ and $N_r$, the number of artificial time steps to make. The numerical experiments in Section~\ref{sec:results} use a mild tapering with $\Delta \tau=0.5$ and $N_r=10$. This tapering process produces a tapered mask that may be applied to the topographic graphic data points. As such, an extended region with ten additional surrounding topographic data points corresponding to the boundary region where tapering is applied is used for quadrilateral grid cells. Applying the tapered mask to non-quadrilateral grid cells is more complex, as the notion of additional surrounding data points does not make much sense, e.g., along the diagonal edge of a triangle embedded in a Cartesian grid. Therefore, the compact tapered mask is initialized by setting values smaller than $10^{-2}$ to zero in non-quadrilateral grid cells.

\begin{figure}[ht]
    \centering
    \includegraphics[width=1.0\textwidth]{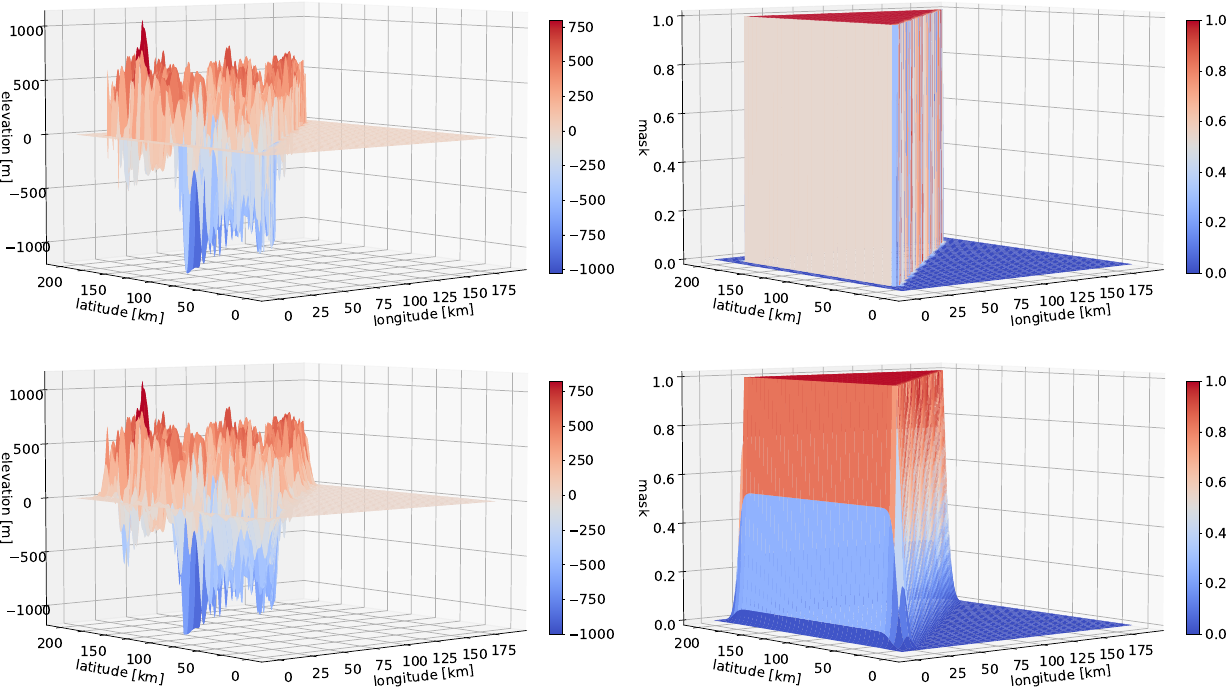}
    \caption{The tapering process applied to the topographic data in grid cell index~158; see also Figure~\ref{fig:delaunay}. Topographic data (top left) and Boolean mask (top right) before the tapering has been applied. Topographic data (bottom left) and tapered mask (bottom right) after tapering. Note that the tapered mask is not Boolean and contains real values in the range $[0,1]$.}
    \label{fig:tapering}
\end{figure}

Figure~\ref{fig:tapering} depicts the result of applying the tapering process to the topographic data in grid cell index~158; see Figure~\ref{fig:delaunay} for the location of this grid cell. Note that $N_r=20$ is applied to the results in this figure. This is to make the effect of tapering more prominent for visual inspection. Before tapering, the topography is sharply truncated at the edges (top left panel) as the mask applied is Boolean (top right panel). This sharp cutoff along the boundary affects the approximated spectrum by introducing high-frequency modes to account for the Gibbs phenomenon. After tapering, the tapered mask (bottom right panel) gradually drops to zero, even at the corners. The mask no longer contains only Boolean values. When applied to the topography (bottom left panel), a gradual decrease to zero is seen along the boundary. The choice of 3D plots here instead of a vertical slice of the domain highlights the effect of tapering data in a non-quadrilateral domain, especially at the corners.

Finally, we note that the application of tapering improves the approximated spectrum. However, these results are not shown here for brevity. For most grid cells, the tapering process leads to a marginal improvement of about 2\% LRE. However, more significant improvement can be seen, e.g., in the grid cells exhibiting considerable overestimation as discussed in Section~\ref{subsec:potential_biases}, where a modest tapering could half the LRE.


\section{Computing the effective pseudo-momentum fluxes}
\label{apx:pmf_computation}
Consider a single monochromatic wave with a given wavenumber $\tilde{k}$ and amplitude $\tilde{a}$. Applying a Fourier transform over a quadrangle encompassing this monochromatic wave produces the stated amplitude and wavenumber. One would expect that the CSAM produces a similar result for each triangle in a Delaunay triangle pair. In such a scenario, the total effective PMF $\mathcal{P}^{\text{eff}}$ would be the average of the fluxes computed in the triangles, i.e.,
\begin{equation}
    \mathcal{P}^{\text{eff}} = \frac{1}{2} \left(  \mathcal{P}^{\text{T1}} + \mathcal{P}^{\text{T2}} \right),
    \label{eqn:pmf_eff_alt}
\end{equation}
which would also be in line with the physical definition of the sum of fluxes in the Delaunay decomposition investigated in Section~\ref{sec:results}.

\begin{figure}
    \centering
    \includegraphics[width=\textwidth]{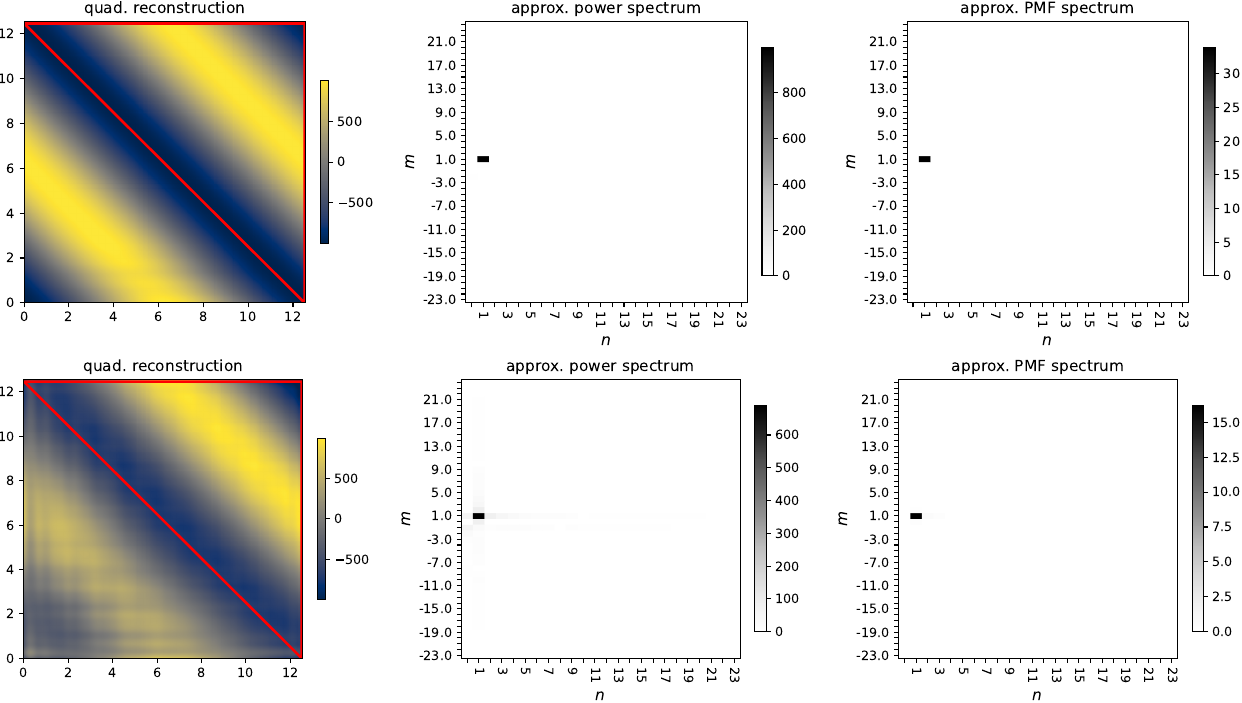}
    \caption{CSAM applied to the data in the upper triangle (enclosed in red) and physically reconstructed over the quadrilateral domain comprising a triangle pair (left panels). The corresponding CSAM approximated spectrum (middle panels) and idealized pseudo-momentum flux spectrum (right panels). A weakly regularized run (top row) and a normally regularized run (bottom row); see text for explanation. As in Section~\ref{subsec:idealized_tests}, the idealized quantities are considered to be dimensionless for brevity.}
    \label{fig:flux_sdy}
\end{figure}

The CSAM only exhibits this behavior if very weak regularization is applied. The regularized CSAM does not reproduce the relation in \eqref{eqn:pmf_eff_alt}, and this observation motivates the definition of $\mathcal{P}^{\text{eff}}$ in \eqref{eqn:pmf_eff}. Specifically, Figure~\ref{fig:flux_sdy} depicts a monochromatic wave with a mode at $(n,m)=(1,1)$. The experimental setup is identical to the one used in Section~\ref{subsec:idealized_tests}, except for $\lambda_{\text{SA}} = 10^{-6}$ for the result in the top row (weakly regularized) and $\lambda_{\text{SA}} = 0.2$ for the bottom row (normally regularized). A Delaunay triangulation is used for the domain decomposition, and only results from the upper triangle half reconstructed over the full quadrilateral domain are shown. Due to the symmetry of the chosen monochromatic wave, the lower triangle half yields results similar to those shown and are therefore omitted.

Compare the left panels in the top and bottom rows. The CSAM is only applied to the data in the upper triangle (enclosed in red). The reconstruction over the quadrangle shows that regularization penalizes the reproduction of features outside the domain of interest. The stronger the regularization, the more the outside features are penalized, decoupling the spectra obtained from the two triangles. Consequently, the magnitude of the approximated amplitude and idealized pseudo-momentum flux differ substantially (compare the color bars between the top and bottom rows for the middle and right panels). Quantitatively, the reference idealized pseudo-momentum flux $\mathcal{P}^{\text{ref}}=33.34$. In comparison, the weakly regularized $\mathcal{P}^{\text{T1}} = 34.06$ and the normally regularized $\mathcal{P}^{\text{T1}} = 17.79$, i.e., approximately half of $\mathcal{P}^{\text{ref}}$, largely representing only the features enclosed in the upper triangle. These results demonstrate that equation \eqref{eqn:pmf_eff_alt} only holds in the limit of no regularization.

However, substantial regularization is unavoidable for complex real-world orography to prevent catastrophic overfitting. Furthermore, unlike monochromatic waves, real-world orographic features beyond the domain of interest have little correlation with the features inside. Therefore, strong enough regularization is favorable to ensure that the approximated spectrum only encodes information from within the domain of interest. These observations motivate the definition in \eqref{eqn:pmf_eff} as a more reasonable approximation for $\mathcal{P}^{\text{eff}}$ as a sum of the fluxes computed from two distinct regularized CSAM spectra.

\acknowledgments
R.C. and U.A. are grateful for the generosity of Eric and Wendy Schmidt through the Schmidt Futures Virtual Earth System Research Institute's DataWave Project. U.A. also thanks the German Research Foundation (DFG) for partial support through CRC 181 ``Energy transfers in Atmosphere and Ocean'' (Project Number 274762653; Projects W01 ``Gravity-wave Parameterization for the Atmosphere'' and S02 ``Improved Parameterizations and Numerics in Climate Models''), and CRC 301 ``TPChange'' (Project Number 428312742; Projects B06 ``Impact of Small-scale Dynamics on UTLS Transport and Mixing'' and B07 ``Impact of Cirrus Clouds on Tropopause Structure'').

The authors thank Young-Ha Kim, Edwin Gerber, François Lott, and Zuzana Procházková for the fruitful discussions that helped improve the development of the constraint spectral approximation method. The authors are also grateful to Axel Seifert for the provision of the MERIT and REMA topographical datasets and Niraj Agarwal for the preliminary studies that led to the development of the method.


%
\bibliography{bib} 
%



\end{document}